\documentclass{aa}  

\usepackage{natbib, multirow}
\usepackage{graphicx}
\usepackage{txfonts}
\usepackage{color}
\usepackage{pdflscape}
\definecolor{aipred}{rgb}{0.63,0.0,0.16}
\definecolor{aipdarkblue}{rgb}{0.0,0.25,0.47}
\usepackage[colorlinks=true,urlcolor=blue,citecolor=aipdarkblue]{hyperref}
\newcommand{\ha}{H$\alpha$}
\newcommand{\hb}{H$\beta$}

\newcommand{\oii}{[\ion{O}{ii}]$\lambda3727$}
\newcommand{\oiii}{[\ion{O}{iii}]$\lambda5007$}
\newcommand{\heii}{\ion{He}{ii}$\,\lambda4686$}

\newcommand{\nii}{[\ion{N}{ii}]$\lambda6584$}

\newcommand{\sii}{[\ion{S}{ii}]$\lambda6716$}

\newcommand{\teff}{$T_{\mathrm{eff}}$}

\begin{document}

   \title{Candidates for the most [\ion{O}{iii}]$\lambda5007$-luminous planetary nebula in the Milky Way}

   \subtitle{I. Integrated light properties of NGC\,6572, NGC\,6884, and M\,1-71}

   \author{Azlizan A. Soemitro \inst{1,2}
          \and
          Martin M. Roth \inst{1,2,3} 
          \and
          Jeremy R. Walsh \inst{4}
          \and
          Ana Monreal-Ibero \inst{5}
          \and
          Brigitte G. A. Pruijt \inst{5}
          \and
          Nicholas Chornay \inst{6,7}
          }

   \institute{Leibniz-Institut für Astrophysik Potsdam (AIP),
              An der Sternwarte 16, 14482 Potsdam, Germany\\
              \email{asoemitro@aip.de}
         \and
            Institut für Physik und Astronomie, Universität Potsdam, Karl-Liebknecht-Str. 24/25, 14476 Potsdam, Germany
        \and
            Deutsches Zentrum für Astrophysik (DZA), Postplatz 1, 02826 Görlitz, Germany
        \and
            European Southern Observatory, Karl-Schwarzschild-Straße 2, 85748, Garching, Germany
         \and
            Leiden Observatory, Leiden University, P.O. Box 9513, 2300 RA Leiden, The Netherlands
        \and
            Department of Astronomy, University of Geneva, Chemin d’Ecogia 16, 1290 Versoix, Switzerland
        \and
            Institute of Astronomy, University of Cambridge, Madingley Road, Cambridge CB3 0HA, UK
             }

   \date{Received ; accepted }

  \abstract
   {}
   {The bright-end cutoff of the planetary nebula luminosity function (PNLF), defined by the \oiii~emission line magnitude as $M_{*} = -4.53$, serves as a standard candle for extragalactic distance measurements. However, the physical properties of planetary nebulae (PNe) at the PNLF cutoff have been studied only in nearby galaxies, where the PNe appear mostly as point sources. Galactic PNLF is further poorly constrained, primarily due to uncertainties in Galactic PN distances. Following a recent Galactic PNLF survey based on Gaia distances, we aim to characterise PN candidates at the Galactic PNLF cutoff. This is the first step toward comparing the extragalactic PNLF cutoff against its Galactic counterpart.}
   {We observed three PN candidates at the Galactic PNLF cutoff using the Potsdam Multi-Aperture Spectrophotometer (PMAS). We determined their chemical abundances and constructed photoionisation models to derive the central star luminosity ($L$) and effective temperature ($T_{\mathrm{eff}}$). Based on their positions in the Hertzsprung–Russell diagram and kinematical ages, we constrained the central star masses and most likely distances consistent with the prediction of stellar evolution models. Using these distances, we determined the absolute magnitude in \oiii~emission line ($M_{5007}$). The PMAS data also allowed us to distinguish the origin of emission lines between the central star and the nebula.}
   {PNe of our sample are located within the top 1 mag of the Galactic PNLF ($-4.20 \lesssim M_{5007} \lesssim-3.60$), with similar nebular properties and a narrow progenitor mass range ($1.30-1.75 \, M_\odot$). The measured extinction is mostly dominated by the foreground Galactic extinction. The trend between the circum-nebular extinction and central star mass of our PNe is in agreement with those derived in the Large Magellanic Cloud (LMC) and M31. All PNe in our sample exhibit weak-emission line star (\textit{wels}) features in their spectra, but we confirmed that some of these lines originate from the nebula.}
   {The PNe around the Galactic PNLF cutoff are similar in terms of their nebular properties and evolutionary stage. Their characteristics are also consistent with the most \oiii-luminous PNe in the LMC and M31. We demonstrate the importance of circum-nebular extinction and the initial-final mass relation (IFMR) to understand the universality of PNLF as a standard candle.}

   \keywords{planetary nebulae: general -- stars: AGB and post-AGB -- galaxies: luminosity function, mass function -- distance scale 
               }

\maketitle

%

\section{Introduction}



Planetary nebulae (PNe) are products of the stellar evolution of low- to intermediate-mass stars ($0.8 - 8.0 \: M_\odot$). Given their various morphologies and chemical abundances, PNe are crucial laboratories for studying stellar evolution and chemical enrichment in the Milky Way and nearby galaxies \citep{2022PASP..134b2001K}. Using \oiii~emission line surveys, we can also construct the planetary nebula luminosity function (PNLF) \citep{1989ApJ...339...39J, 1989ApJ...339...53C}. In recent years, the PNLF has gained added interest for its use as extragalactic distance indicator \citep{2017ApJ...834..174K, 2021A&A...653A.167S, 2021ApJ...916...21R, 2022MNRAS.511.6087S, 2023A&A...671A.142S, 2024ApJS..271...40J, 2025A&A...700A.125C, 2025A&A...704A.303S} and as a tracer of stellar populations in galaxies \citep{2013A&A...558A..42L, 2015A&A...579A.135L, 2017A&A...603A.104H, 2020A&A...642A..46H, 2025FrASS..1273373H, 2026MNRAS.545f2036E}. 

The PNLF exhibits an absolute bright-end cutoff magnitude of $M_* = -4.53 \pm 0.06$ \citep{2012Ap&SS.341..151C}; this is a standard candle for measuring extragalactic distances that has been universally demonstrated across different Hubble types in comparison to other techniques such as Cepheids and tip of the red giant branch. The distance reach of the PNLF has been significantly improved from $\sim 15$ Mpc in the early 2010s to $\sim 40$ Mpc in the last few years \citep{2021ApJ...916...21R, 2022MNRAS.511.6087S, 2024ApJS..271...40J}. The theoretical foundation of the universal bright-end cutoff has also seen a major advance. Recent post asymptotic giant branch (post-AGB) models \citep{2016A&A...588A..25M} and related PN simulations \citep{2018NatAs...2..580G, 2019ApJ...887...65V, 2025A&A...699A.371V}, including the studies of initial-final mass relation \citep[IFMR,][]{2018ApJ...866...21C, 2024MNRAS.527.3602C} and PN circum-nebular extinction, \citep{2025ApJ...983..129J, 2025FrASS..1209047V} have helped to close the gap between the theory and observation. However, until now, the physical understanding of the PNLF bright-end cutoff has been based on external galaxies, where the PNe appear mostly as point sources. As they are likely spatially resolved, studies of PNe at the bright-end of the Galactic PNLF cutoff are highly desirable to investigate the physical foundations of the PNLF as a standard candle.

The main challenge in deriving the Galactic PNLF has been the notoriously uncertain distances to Galactic PNe. Reliable distances could be obtained for only a small number of objects, making population studies extremely difficult. Fortunately, this situation is changing. \citet{2016MNRAS.455.1459F} derived robust statistical distances for a large number of PNe using the H$\alpha$ surface brightness–radius relation. Moreover, the \textit{Gaia} space mission \citep{2016A&A...595A...1G} has, for the first time, provided geometrical parallax distances for a significant number of PNe \citep{2018A&A...616L...2K, 2021A&A...656A..51G, 2021A&A...656A.110C}. This enables investigations of the Galactic PNLF using large samples in the solar neighbourhood. A preliminary study of the local Galactic PNLF was carried out by \citet{Chornay_Walton_Jones_Boffin_2023} using narrowband \oiii~imaging and \textit{Gaia} distances from \citet{2021A&A...656A.110C} (see their Fig. 2 for details). Based on this work, it is now possible to identify the PNe located at the bright-end cutoff of the Galactic PNLF. 


In this paper, we aim to investigate the physical properties of the most \oiii-luminous PN candidates in the Milky Way. Our goal is to analyse their nebular properties, including electron temperature ($T_e$), electron density ($N_e$), extinction ($c(\mathrm{H}\beta)$), and chemical abundances. This enables the construction of photoionisation models to derive the central star luminosity ($L$) and effective temperature ($T_{\mathrm{eff}}$). We also estimate the initial and final masses of the central stars using recent post-AGB evolutionary models and various initial–final mass relations (IFMRs). Furthermore, we independently determine their absolute \oiii~magnitudes ($M_{5007}$) and infer their evolutionary stages. The classification of the central stars and its implications for their evolutionary histories are also discussed.

This paper is structured as follows. Sect. \ref{sec:data} describes the data for our analysis. Sect. \ref{sec:nebular_diag} explains the methodology of the nebular diagnostics. Sect. \ref{sec:pn_distance} elaborates the refinement details on the literature distance, which is achieved from photoionisation models, post-AGB models, and kinematic ages. Sect. \ref{sec:m5007} explains the determination of $M_{5007}$ and the correction of the foreground Galactic extinction. The results on the nebular properties, the central star properties, and the $M_{5007}$ values are presented in Sect. \ref{sec:result}. The implications for the physical foundation of the PNLF bright-end cutoff are discussed in Sect. \ref{sec:discuss}. Finally, we conclude our paper in Sect. \ref{sec:conclud}.







\section{Data} \label{sec:data}

\subsection{Sample description}

The study by \citet{Chornay_Walton_Jones_Boffin_2023} currently provides the most statistically significant Galactic PNLF, based on \textit{Gaia} distances. We selected three of their most \oiii-luminous PNe as test cases for a detailed physical investigation: NGC\,6572, NGC\,6884, and M\,1-71. 

NGC\,6572 is a compact, bipolar, and high excitation PN in the northern sky. It has an angular size of 13\arcsec$\times$15\arcsec \citep{2016MNRAS.455.1459F}. Currently, there are several distances to the PN, ranging between $\sim1.2-1.8$ kpc \citep{1995AJ....109.2600H, 2016MNRAS.455.1459F, 2021A&A...656A.110C}. Previous spectroscopic studies and modelling suggest a central star progenitor of $\sim1.00-1.25 \,M_\odot$, classified as a weak emission central star (\textit{wels}) \citep{1994MNRAS.269..975H, 2023MNRAS.524.1547B} . Although the line intensities of the nebula appear to be constant, long-term observations showed a hint of a 70-year cycle of the central star spectrum \citep{1994MNRAS.269..975H}.

NGC\,6884 is also a compact, high excitation PN, with an angular size of 7.5\arcsec$\times$7.0\arcsec \citep{2016MNRAS.455.1459F}. The Hong kong/AAO/Strasbourg PN catalogue \citep[HASH,][]{2016JPhCS.728c2008P} classified its morphology as elliptical. Distance estimates range between 2.2 and 5.0 kpc \citep{2002AJ....123.2666P, 2016MNRAS.455.1459F, 2021A&A...656A.110C}. Multi-wavelength studies revealed a complex set of low ionisation structure, as observed in the emission line of \nii, and bipolar outflows \citep{1997ApJS..108..503H, 1999AJ....117.1421M}. The central star has an uncertain classification of perhaps [WN] \citep{1997ApJS..108..503H}. Its progenitor was thought to be a $\sim1\,M_\odot$ star.

M\,1-71 is another compact and high excitation PN with an angular size of 3.7\arcsec$\times$6.0\arcsec \citep{2016MNRAS.455.1459F}. Compared to the other two PNe, it is the least studied one. It is estimated to have a distance between 2.80 and 3.30 kpc \citep{2016MNRAS.455.1459F, 2021A&A...656A.110C}. \citet{2005A&A...436..967W} conducted a spectrophotometric study and derived some chemical abundances; they did not found peculiarities. This PN was also observed as a calibration target during the commissioning and demonstration of the SITELLE instrument \citep{2019MNRAS.485.3930D}. They revealed a bipolar core from the emission line of \nii. The central star was classified as \textit{wels}, although its parameters have not been determined \citep{1998A&A...329L...9P}.

\subsection{Observations}

We observed NGC\,6572, NGC\,6884, and M\,1-71 using the Potsdam Multi-Aperture Spectrophotometer (PMAS), which is an integral field spectrograph mounted on the Calar Alto 3.5-meter telescope \citep{2005PASP..117..620R}. We used the lens array setup with 0.5\arcsec/pixel sampling that provides an 8\arcsec $\times$ 8\arcsec field-of-view (FOV). We employed the V600 grating with a grating angle of 144.5 degrees ($\lambda \sim 3600-6800\:\AA ; \: R\sim1500$) for NGC\,6572. For NGC\,6884 and M\,1-71, we used V500 grating with a grating rotator position of 143.5 units ($\lambda \sim 3600-7500\:\AA ; \: R\sim1300$). The observation details are summarised in Table \ref{tab:dataset overview}. We created a mosaic for NGC\,6572, while the other two objects were covered with single-pointing observations. HgNe arc lamp exposures for the wavelength calibration were taken before and after each object. Two standard stars were observed per night, at the beginning and the end of the night. We found that the flux calibration is accurate within at least $5\%$; the details are explained in Appendix \ref{app:flux_cal}. We also observed offset sky exposure of 300 seconds for each target. 

\begin{table*}
\caption{Observation overview.}
\label{tab:dataset overview}
\centering
\begin{tabular}{c c c c c c}
\hline
\hline
Name & Angular size\tablefootmark{a} & Observing date & Pointing & Exposure Time & Seeing FWHM\tablefootmark{b} \\
\hline
\hline
 &  &  & P1 & $6 \times5$s & \\
 &  &  & P2 & $6 \times5$s & \\
 &  &  & P3 & $6 \times5$s & \\
 &  &  & P4 & $6 \times20$s & \\
 &  &  & P5 & $6 \times30$s & \\
NGC\,6572 & 13\arcsec $\times$ 15\arcsec & 2023-09-12 & P6 & $6 \times30$s & 1\farcs3 -- 1\farcs5\\
 &  &  & P7 & $6 \times30$s & \\
 &  &  & P8 & $6 \times30$s & \\
 &  &  & P9 & $6 \times30$s & \\
 &  &  & P10 & $6 \times30$s &\\
 &  &  & P11 & $6 \times30$s &\\

\hline
NGC\,6884 & 7.0\arcsec $\times$ 7.5\arcsec  & 2024-05-06 & P1 & $3 \times30$s& 1\farcs5\\
\hline
M\,1-71 & 3.7\arcsec $\times$ 6.0\arcsec & 2024-05-06 & P1 & $3 \times180$s& 1\farcs9\\
\hline
\end{tabular}
\tablefoot{
\tablefoottext{a}{Based on the \ha~surface brightness catalogue of \citet{2016MNRAS.455.1459F}.}
\tablefoottext{b}{Measured using the guiding camera with V-band filter at the beginning and the end of the exposure.}
}
\end{table*}

\subsection{Data reduction and post-processing} \label{sec:data_red}

The data were reduced, wavelength-, and flux-calibrated using the \textsc{p3d} software \citep{2010A&A...515A..35S}. We found that some spaxels were giving unphysical values. We speculated that this was caused by ageing of the associated optical fibre interface material and vignetting. 

By comparing the target exposure and offset sky exposure, we found that the problematic spaxels between the two were consistent, although the values can be discrepant. To minimise this effect, we traced the wavelength-dependent bad spaxels in the offset sky exposure and masked everything with a flux larger than $5 \times 10^{-15}$ erg s$^{-1}$ cm$^{-2}$ \AA$^{-1}$. Then, we applied the same mask to our target exposures. The cleaning procedure was done in \textsc{Python}. The vignetting affected the outermost columns and rows of spaxels in the FOV. The effect was especially bad on the bluest and reddest part of the spectrum. To overcome this, we additionally masked the outermost two columns and rows, giving an effective FOV of 6\arcsec $\times$ 6\arcsec. The full cleaning procedure is illustrated in Appendix \ref{app:mosaic}. Afterward, for each pointing, we applied an atmospheric refraction correction and transformed it into a datacube using built-in procedures in \textsc{p3d}.  

To create the mosaic of NGC\,6572, we used the \textsc{astropy.reproject.mosaicking} package\footnote{https://reproject.readthedocs.io/en/stable/}. First, we used \textsc{find\_optimal\_celestial} to read the coordinate of each pointing, re-gridded, and defined a single, generalised coordinate system of all pointings. Then, using \textsc{reproject\_and\_coadd}, individual pixels from the images were projected and interpolated to construct a single mosaic image. Due to the limited FOV of our individual pointing, we were unable to make a better astrometric alignment using background stars. Therefore, we qualitatively compared our mosaic with an image of NGC\,6572 observed with the Hubble Space Telescope (HST), which was obtained with the WFPC2 instrument and F502N filter (Program ID: 9839). Since our analysis was focused on the integrated light, with mostly comparing line ratios, we found that our mosaicking technique was sufficient. The final mosaic is also presented in Appendix \ref{app:mosaic}.

\section{Nebular diagnostics} \label{sec:nebular_diag}

To obtain the PN spectra, we integrated the whole FOV. Then, we fitted a Gaussian to the emission lines using the \textsc{LMFIT} package \citep{2020zndo...3814709N} in \textsc{Python}. The outcomes of the fit were the line flux, the centre wavelength, the full-width-half-maximum (FWHM), and the line flux uncertainty. The final line flux uncertainty was calculated by convolving the fit error and the flux calibration error of 5\%. For nebular diagnostics, we employed \textsc{PyNeb} \citep{2015A&A...573A..42L}. We measured each diagnostics using the Monte Carlo approach with 1000 iterations, assuming a Gaussian distribution of the uncertainty. The final result, lower, and upper uncertainties were represented by the median, the 16$^\mathrm{th}$ percentile, and the 84$^\mathrm{th}$ percentile, respectively. The specific lines and diagnostic purposes are discussed in the following subsections. The references for each atomic line emissivity are tabulated in Appendix \ref{app:line_emissivity}. The nebular diagnostics are our basis to create photoionisation models for our PNe; this is described in Sect. \ref{sec:cloudy}. 

\subsection{Extinction, electron temperature, and electron density} \label{sec:ext,te,ne}

We calculated the extinction ($c(\mathrm{H}\beta)$), electron temperature ($T_e$), and electron density ($N_e$) iteratively. First, we measured the extinction using the Balmer line ratio of \ha~and \hb, assuming a standard assumption for case B recombination of $T_e = 10000$ K and $N_e = 1000$ cm$^{-2}$ and the extinction law from \citet{1989ApJ...345..245C} with $R_{V} = 3.1$. Using this value, we calculated the first iteration of $T_e$ and $N_e$ using the extinction-corrected line fluxes. For low ionisation components we derived the parameters of $T_e$[\ion{N}{ii}]$\lambda5755,6584$ and $N_e$[\ion{S}{ii}]$\lambda6716,6731$. For higher ionisation components, we calculated $T_e$[\ion{O}{iii}]$\lambda4363,5007$, $N_e$[\ion{Cl}{iii}]$\lambda5518,5538$, and $N_e$[\ion{Ar}{iv}]$\lambda4712,4740$. Then, we assumed $T_e$[\ion{N}{ii}] and $N_e$[\ion{S}{ii}] to measure the extinction again and repeated every $T_e$ and $N_e$ calculations. Note that the [\ion{Ar}{iv}]$\lambda4712$ emission line might be blended with \ion{He}{I}$\lambda4713$, which is unresolved in our spectra. The values typically converge after two or three iterations. We used the final extinction value to de-redden all of the emission line fluxes for the abundance calculation, assuming the same extinction law. 

\subsection{Ionic and elemental abundances} \label{sec:chem_abund}
\subsubsection{Helium}

To determine the ionic abundance of He$^+$, we used the emission lines of \ion{He}{i}$\,\lambda5876$ and \ion{He}{i}$\,\lambda6678$. Note that the \ion{He}{i}$\,\lambda6678$ line is a singlet, while the \ion{He}{i}$\,\lambda5876$ one is a triplet. Triplet lines are more sensitive to radiative transfer effects, and therefore they must be corrected from this effect before using them for abundance determination \citep{1968ApJ...151..497R, 2013A&A...553A..57M}. The correction factor $f_\tau(\lambda)$ is a function of optical depth at $\lambda3889$. To calculate the optical depth $\tau(\lambda3889)$, we followed the procedure described in \citet[][Sect. 3.2.3]{2013A&A...553A..57M} and \citet[][Sect. 7.1]{2018A&A...620A.169W}. Assuming a spherically symmetric expanding nebula, we fitted the $\tau$ values derived by \citet[][Table 3]{1968ApJ...151..497R} to the functional form of $f_\tau(\lambda)=1 + a\tau^b$.

We adopted the values for a ratio between the expansion and thermal velocity $V(R) / V(TH) = 5$ and $T_e = 10000$ K, which was the largest ratio calculated by \citet{1968ApJ...151..497R}; our PNe has the ratio of $\sim7-17$. After the correction, we calculated the ionic abundance of He$^+$ assuming $T_e$[\ion{N}{ii}] and $N_e$[\ion{S}{ii}]. We adopted the average abundance from the lines as our final result.

For He$^{++}$ abundance, we employed the emission line of \ion{He}{ii}$\,\lambda4686$. This was only applicable for NGC\,6884 and NGC\,6572. Although we detected a faint hint of the line in M\,1-71, due to a low signal, we did not attempt any measurement. For the abundance calculation, we assumed $T_e$[\ion{O}{iii}] and $N_e$[\ion{Ar}{iv}]. 

The elemental abundance of helium was derived without any ionisation correction factor (ICF). Based on the oxygen-line cutoff-criterion scheme \citep[][BSJ]{2000ApJ...536..773B, 2002A&A...381..361S}, the correction is not necessary when the line ratio of \oiii/\hb~is $\gtrsim8$; our PNe have the ratio \oiii/\hb~$\gtrsim12$. Moreover, since we do not have any tracer for He$^0$, we adopt the assumption that its contribution was negligible. The neutral helium can be estimated from photoionisation models \citep{2018A&A...620A.169W}. Based on the modelling explained in Sect. \ref{sec:cloudy}, the models of our PNe consisted of $\sim$1 \% of He$^0$. The total helium abundance was calculated with a summation of He$^+$ and He$^{++}$.

\subsubsection{Oxygen}

We measured the ionic abundances of O$^0$, O$^+$, and O$^{++}$. For the O$^0$ abundance, we employed the emission lines of [\ion{O}{i}]$\lambda6300$ and [\ion{O}{i}]$\lambda6363$. To derive the O$^+$ abundance of NGC\,6884 and M\,1-71, we considered the emission lines of [\ion{O}{ii}]$\lambda3727$, [\ion{O}{ii}]$\lambda7220$, and [\ion{O}{ii}]$\lambda7230$. For NGC\,6572, only [\ion{O}{ii}]$\lambda3727$ was available within the wavelength range. This [\ion{O}{ii}] line is actually an unresolved doublet of [\ion{O}{ii}]$\lambda3727,29$; we fitted the emission line with a single Gaussian, but we considered them as doublets in \textsc{PyNeb}. We assumed $T_e$[\ion{N}{ii}] and $N_e$[\ion{S}{ii}] for the abundance calculation of the O$^0$ and the O$^+$. Lastly, the abundance measurement of O$^{++}$ were carried out using the [\ion{O}{iii}]$\lambda5007$ emission line for all PNe. $T_e$[\ion{O}{iii}] and $N_e$[\ion{Ar}{iv}] were assumed for this calculation. 

The total abundance of oxygen was calculated by adding O$^+$ and O$^{++}$. The O$^0$ abundance was excluded because it mostly comes from photodissociation regions \citep{1991ApJ...377..210R, 1995MNRAS.273...47L, 2018A&A...620A.169W}. Our modelling confirmed that the O$^0$ contribution was rather low; $\lesssim$3 \% for all PNe. Then, to take account of all ionic stages above O$^{++}$, we applied the ICF. Several ICF schemes are available in the literature, e.g., \citet[][KB94]{1994MNRAS.271..257K} and \citet[][DI14]{2014MNRAS.440..536D}. We employed both to see how they compare to each other. The ICF uncertainties of KB94 were propagated from the line flux error, while DI14 provided their own recipe for the uncertainties. The uncertainty calculation for other elements followed the same procedure.  

\subsubsection{Nitrogen, sulphur, argon, neon, and chlorine}

With our spectra, it was only possible to derive the ionic abundance of N$^+$. We used the emission lines of [\ion{N}{ii}]$\lambda5755$, [\ion{N}{ii}]$\lambda6548$, and [\ion{N}{ii}]$\lambda6584$. The $T_e$[\ion{N}{ii}] and $N_e$[\ion{S}{ii}] were adopted for the calculation. To derive the elemental nitrogen abundance, we applied the ICF schemes from KB94 and DI14.

There were two sulphur transitions in all of our PNe; S$^+$ traced using [\ion{S}{ii}]$\lambda6716,31$ emission lines and S$^{++}$ traced using the emission line of [\ion{S}{iii}]$\lambda6312$. We used the values of $T_e$[\ion{N}{ii}],$N_e$[\ion{S}{ii}] to derive the S$^+$ abundance and $T_e$[\ion{O}{iii}],$N_e$[\ion{Ar}{iv}] for S$^{++}$ abundance. We calculated the ICF using the schemes of KB94 and DI14 to derive the elemental abundance of sulphur. 

For NGC\,6884 and M\,1-71, we derived the ionic abundances of Ar$^{++}$ and Ar$^{+++}$ from the emission lines of [\ion{Ar}{iii}]$\lambda7135$ and [\ion{Ar}{iv}]$\lambda4712,40$, respectively. For NGC\,6572, however, only the [\ion{Ar}{iv}] lines were available and therefore only Ar$^{+++}$ abundance was derived. We adopted $T_e$[\ion{O}{iii}] and $N_e$[\ion{Ar}{iv}] values for both ionic abundance calculation. To derive the elemental argon abundance of NGC\,6884 and M\,1-71, we used the KB94 ICF scheme that took account of Ar$^{++}$ and Ar$^{+++}$ and the DI14 scheme that only need the Ar$^{++}$ abundance as input. For NGC\,6572, we used the ICF from KB94 that only requires the ionic abundance of Ar$^{+++}$. 

For neon abundance, it was only possible to derive the ionic abundance of Ne$^{++}$. We used the emission line of [\ion{Ne}{iii}]$\lambda3869$, assuming the values of $T_e$[\ion{O}{iii}] and $N_e$[\ion{Ar}{iv}]. The ICF was derived using the schemes from KB94 and DI14 that only considered the Ne$^{++}$ ionic abundance. 

The chlorine abundance was derived using the doublet of [\ion{Cl}{iii}]$\lambda5518,38$. The values of $T_e$[\ion{O}{iii}] and $N_e$[\ion{Ar}{iv}] were assumed for the calculation. The ICF was determined using the DI14 scheme. For M\,1-71, the [\ion{Cl}{iii}]$\lambda5538$ line was the only one detected.




\subsection{Excitation class}

We determined the excitation class \citep{1990ApJ...357..140D, 2010PASA...27..187R} for our PNe, which is defined as  

\begin{equation}
    EC_{\mathrm{low}} = 0.45 \: \bigg[\frac{F(\lambda5007)}{\mathrm{H}\beta}\bigg]
\end{equation}

\begin{equation}
    EC_{\mathrm{high}} = 5.45 \: \bigg[\frac{F(\lambda4686)}{\mathrm{H}\beta} + \mathrm{log_{10}}\frac{F(\lambda4959 + \lambda5007)}{\mathrm{H}\beta}\bigg]
\end{equation}

\noindent
We used $E_{\mathrm{low}}$ to determine the excitation class of NGC\,6572 and M\,1-71. Although we detected a weak emission line of \heii~in both PNe, we speculate that the \heii~emission in these PNe might come from the central star. This is further discussed in Sect. \ref{sec:cs_class}. For NGC\,6884, we confirmed the nebular nature of \heii~from the emission line map and therefore we employed $E_{\mathrm{high}}$. We used the excitation class to infer the evolutionary phase of our PNe and to see whether they can be associated with the bright-end cutoff of the PNLF \citep{2007A&A...473..467S}. 

\section{PN distances} \label{sec:pn_distance}

Distances of Galactic PNe are uncertain up to a factor of two or more. To assess the most likely true distance, we considered several literature values obtained with different methods; \textit{Gaia}-based distances from \citet[][CW21]{2021A&A...656A.110C} that was used for the preliminary Galactic PNLF \citep{Chornay_Walton_Jones_Boffin_2023}, H$\alpha$ surface brightness statistical distances \citep[][F16]{2016MNRAS.455.1459F}, and expansion parallax distances (\citealt{1995AJ....109.2600H} (H95) for NGC\,6572; \citealt{2002AJ....123.2666P} (P02) for NGC\,6884). There is no expansion parallax distance in the literature for M\,1-71. The distances of \citet{2021A&A...656A.110C} were determined in two ways; a parallax measurement and a statistical approach that used the parallax as a prior. For NGC\,6572, the first option was available and we selected it. For the other two objects, only the latter was available. The distances to each PNe are shown in Table \ref{tab:distances}.

\begin{table}
\caption{PN distances with their corresponding $L_*$ and $t_\mathrm{kin}$.}
\label{tab:distances}
\centering
\begin{tabular}{c c c c}
\hline
\hline
\multirow{2}{*}{Reference} & \multicolumn{3}{c}{NGC\,6572} \\
    & $d$ [kpc]& log $L_*$ [$L_\odot$] & $t_{kin}$ [yr] \\ 

\hline
\hline
H95 & 1.18$\pm$0.34 & 3.37$_{-0.35}^{+0.25}$ & 2821$_{-1006}^{+1434}$ \\
\hline
F16 & 1.46$\pm$0.42  & 3.55$_{-0.33}^{+0.28}$ & 3512$_{-1223}^{+1794}$\\
\hline
CW21 & 1.82$_{-0.16}^{+0.18}$ & 3.74$_{-0.15}^{+0.16}$ & 4344$_{-1006}^{+1631}$\\
\hline
This work & 1.83$_{-0.25}^{+0.38}$ & 3.76$_{-0.19}^{+0.22}$ & 4434$_{-1147}^{+1935}$ \\
\hline
\hline

\multirow{2}{*}{} & \multicolumn{3}{c}{NGC\,6884} \\
\hline
\hline
P02 & 2.20$\pm$0.80 & 2.83$_{-0.41}^{+0.29}$ & 2304$_{-792}^{+814}$\\
\hline
F16 & 3.22$\pm$0.92 & 3.15$_{-0.33}^{+0.24}$ & 3360$_{-1002}^{+968}$ \\
\hline
CW21 & 4.94$_{-0.66}^{+0.71}$ & 3.53$_{-0.16}^{+0.16}$ & 5137$_{-644}^{+742}$\\
\hline
This work & 5.47$_{-0.93}^{+1.04}$ & 3.62$_{-0.21}^{+0.18}$ & 5678$_{-1001}^{+1057}$\\
\hline
\hline

\multirow{2}{*}{} & \multicolumn{3}{c}{M\,1-71} \\
\hline
\hline
F16 & 2.88$\pm$0.91 & 3.43$_{-0.36}^{+0.27}$ & 883$_{-290}^{+287}$\\
\hline
CW21 & 3.27$_{-0.76}^{+0.82}$ & 3.57$_{-0.29}^{+0.22}$ & 1005$_{-220}^{+272}$\\
\hline
This work & 4.35$_{-0.63}^{+0.77}$ & 3.81$_{-0.19}^{+0.19}$ & 1356$_{-210}^{+228}$\\
\hline
\hline

\end{tabular}
\tablefoot{H95: \citet{1995AJ....109.2600H}, P02: \citet{2002AJ....123.2666P}, F16: \citet{2016MNRAS.455.1459F}, CW21: \citet{2021A&A...656A.110C}}
\end{table}

The distance is crucial to determine the central star luminosity ($L_*$) and the absolute magnitude $M_{5007}$. Since the variation of literature distances is large, our result will be highly dependant on the distance choice. To obtain a better distance constraint, we assessed the age of the PNe. Unlike extragalactic PNe, our PNe are spatially resolved and it is possible to estimate the kinematical age ($t_{\mathrm{kin}}$), which is also a function of distance. Testing the consistency between the $L_*$, $t_{\mathrm{kin}}$, and the theoretical stellar age from the post-AGB tracks ($t_{\mathrm{tracks}}$) should provide additional constrains that are needed for our analysis. For the comparison with the stellar tracks, we determined the central star effective temperature ($T_\mathrm{eff}$) through photoionisation modelling. The technical details are described in the subsections below.  

\subsection{Central star luminosity}

To derive $L_*$, we used the de-reddened H$\beta$ flux to calculate the H$\beta$ luminosity ($L_{\mathrm{H}\beta}$), assuming a distance. We used the Monte Carlo approach with 1000 iterations, considering the uncertainties of the extinction and distance. Then we determined the photon production rate, $Q(\mathrm{H})$, using the equation 

\begin{equation}
    Q(\mathrm{H}) = \frac{\alpha_B}{\epsilon_{\mathrm{H}\beta}} \: L_{\mathrm{H}\beta},
\end{equation}

\noindent
where $\alpha_B$ is the total recombination coefficient assuming the case-B recombination theory and $\epsilon_{\mathrm{H}\beta}$ is the line emissivity of H$\beta$. Both parameters are a function of $T_e$ and $N_e$, that were calculated using \textsc{PyNeb}. The uncertainty of $Q(\mathrm{H})$ was propagated from $L_{\mathrm{H}\beta}$. Afterward, we adopted the spectral energy distribution (SED) of Rauch H-Ni model atmospheres \citep{2003A&A...403..709R} assuming a constant log $g$ = 6.5 to determine the ionising surface flux

\begin{equation}
    q_{\mathrm{H}} = \int_0^{\lambda_L} \frac{F(\lambda)}{hc} \lambda \,d\lambda,
\end{equation}

\noindent
where $\lambda_L$ is the Lyman edge at 912\AA, $h$ is the Stefan-Boltzmann constant, and $c$ is the speed of light. These Rauch models were adopted to study the \oiii-luminous PNe in M31 \citep{2022A&A...657A..71G}. We also calculated the bolometric flux 

\begin{equation}
    F_\mathrm{bol} = \int_0^\infty F(\lambda) \, d\lambda.
\end{equation}

\noindent
For the first iteration, we calculated the luminosity by assuming $T_\mathrm{eff} = 100\,000$ K for all PNe to obtain the initial $L_*$ for photoionisation modelling. This value is later revised with the $T_\mathrm{eff}$ output from the final models. Regardless, we found that the effect of $T_\mathrm{eff}$ to be negligible for this analysis. Finally, we can determine the $L_*$ using

\begin{equation}
    L_* = Q(\mathrm{H}) \: \frac{F_{\mathrm{bol}}}{q_\mathrm{H}}.
\end{equation}

\noindent
The uncertainty for $L_*$ was propagated from $Q(\mathrm{H})$. The values are presented in Table \ref{tab:distances}.

\begin{figure*}
   \centering
   \includegraphics[width=0.9\linewidth]{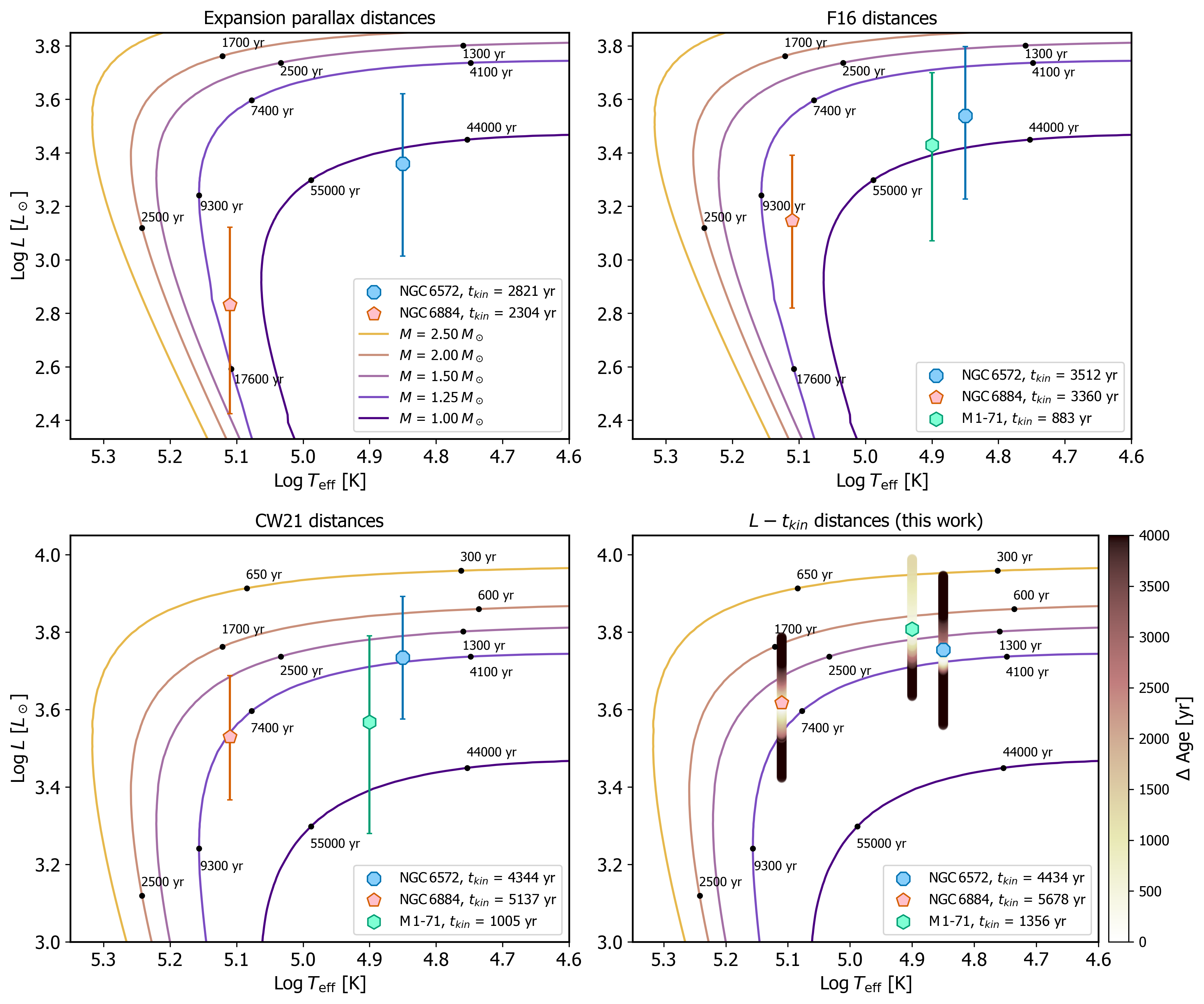}
   \caption{Comparison of PN locations in the HR-diagram when adopting the expansion parallax distances of H95 and P02 (top left), \ha~surface brightness distances of F16 (top right), Gaia distances of CW21 (bottom left) and $L_*-t_\mathrm{kin}$ distances (bottom right). NGC\,6572, NGC\,6884, and M\,1-71 is marked with blue octagon, green hexagon, and orange pentagon. The stellar tracks are post-AGB models from \citet{2016A&A...588A..25M} with $Z_\odot=0.01$. The shaded regions in the bottom right panel indicate the $L_*-t_\mathrm{kin}$ distance uncertainties, with the colour indicates the age difference between $t_\mathrm{kin}$ and $t_\mathrm{tracks}$ for each PN.}
   \label{fig:distance_tkin}
\end{figure*}

\subsection{Kinematical age}

The kinematical age can be determined using 

\begin{equation}
    t_{\mathrm{kin}} = \frac{\theta_r \: d}{v_\mathrm{exp}},
\end{equation}

\noindent
where $\theta_r$ is the angular radius of the PN, $d$ is the adopted distance, and $v_\mathrm{exp}$ is the PN expansion velocity. We calculated the $\theta_r$ values from the angular diameters of \citet{2016MNRAS.455.1459F}, as listed in Table \ref{sec:data}. We calculated an average radius, assuming a circular geometry. The literature values of $v_\mathrm{exp}$ that we adopted for each PN is presented in Table \ref{tab:expansion_velocity}. The $v_\mathrm{exp}$ value for M\,1-71 is larger than NGC\,6572 and NGC\,6884 by a factor of two. However, its value is still within the expected $v_\mathrm{exp}$ of PNe, when compared to the study of \citet[][see their Fig. 6 for details]{2013A&A...558A..78J}. The $t_{\mathrm{kin}}$ error was propagated from the errors of the distance and the expansion velocity. In this case, the distance error was the dominating source of uncertainty.

\begin{table}
\caption{Literature $v_\mathrm{exp}$ of our PN sample.}
\label{tab:expansion_velocity}
\centering
\begin{tabular}{c c c}
\hline
\hline
PN & $v_\mathrm{exp}$ [km/s] & Reference \\
\hline
\hline
NGC\,6572 & 14.0$\pm$4.0 (\oiii) & \citet{1995AJ....109.2600H} \\
\hline
NGC\,6884 & 16.6$\pm$0.4 (\oiii) & \citet{2002AJ....123.2666P} \\
\hline
M\,1-71 & 38.5$\pm$1.0 (\ha) & \citet{2019MNRAS.485.3930D} \\

\hline
\hline
\end{tabular}
\end{table}

This simplistic approach had some caveats as it assumed a spherical expansion with a constant velocity and typically based on a single line. Firstly, the constant velocity assumption may not be exactly true, as PNe are predicted to be accelerated by a factor of two since the AGB phase \citep{2000A&A...358.1058G}. Furthermore, based on 1D-hydrodynamical modelling, \citet{2013A&A...558A..78J} have explored the true expansion velocity throughout the PN evolution. They found that the \oiii~velocity tends to be lower than the \nii~velocity; this was especially true when the PN had a considerable shock component. To obtain a true velocity from the \oiii~line, \citet{2013A&A...558A..78J} suggested a correction factor of $\sim$1.25. The \oiii~and \ha~are typically observed with similar velocity fields \citep{2018A&A...620A.169W, 2020A&A...634A..47M}, so the same correction factor can be applied. However, at this stage of analysis, we did not to apply any correction factor because uncertainty contribution of $v_\mathrm{exp}$ to the $t_\mathrm{kin}$ value was insignificant compared to the distance uncertainty. 

For each distance, we calculated the $t_\mathrm{kin}$ and show them in Table \ref{tab:distances}. We compared the $L_*$ and $t_{\mathrm{kin}}$ values with the expectation from post-AGB models \citep{2016A&A...588A..25M} in Fig. \ref{fig:distance_tkin}; the \teff~ derivation is discussed in Sect. \ref{sec:cloudy} and Sect. \ref{sec:ltkin_distance}. We found that the adopted distance has a significant impact on the agreement between the measured $t_{\mathrm{kin}}$ and the expected $t_{\mathrm{tracks}}$. This implies that we can derive the most likely distance, when the difference between $t_{\mathrm{kin}}$ and $t_{\mathrm{tracks}}$ is at its minimum, for a given central star $L_*$ and \teff.

\subsection{Photoionisation models} \label{sec:cloudy}

To determine $T_{\mathrm{eff}}$, we performed photoionisation modelling using \textsc{Cloudy} \citep[C23.01,][]{2023RNAAS...7..246G} through \textsc{pyCloudy} interface \citep{2013ascl.soft04020M}. Again, we adopted the Rauch H-Ni model atmospheres \citep{2003A&A...403..709R} with log g = 6.5. In the first model iteration, we assumed our measured chemical abundances and $L_*$ derived using CW21 distances; their $L_*$ and $t_{\mathrm{kin}}$ values had the best agreement with the post-AGB stellar tracks \citep{2016A&A...588A..25M}. We ran a grid of models assuming a spherical geometry with a constant density with varying effective temperature with the range of log $T_{\mathrm{eff}} = 4.8 -5.2$. A constant density was assumed for previous modelling on extragalactic PNe \citep{2012ApJ...753...12K, 2022A&A...657A..71G, 2025ApJ...983..129J}. To have a consistent comparison between our work and theirs, we adopted the same density assumption. We employed the density of $N_e$[\ion{Cl}{iii}] when available, otherwise we used $N_e$[\ion{Ar}{iv}]. Since our sample consists of  medium- to high excitation PNe ($EC > 5$, see Sect. \ref{sec:oiii_evolution} for details), the nebulae are dominated by the medium- to high ionisation zones. Therefore, our selection of higher ionisation density tracer is more appropriate than a lower ionisation tracer such as $N_e$[\ion{S}{ii}]. Moreover, because we did not have any tracers of carbon abundance, we assumed an abundance ratio of C$/$O = 1.0, which was shown to be an acceptable compromise \citep{2022A&A...657A..71G}. We noted that the spherical geometry might not be the best assumption for our PNe. However, the simple geometry was typical in analysis of extragalactic PNe or PNLF simulation. 

To select the best model, we compared the relative line strength to \hb~between the model and the observation. Moreover, we considered the emission line ratios described in \citet[][Table 6]{2012ApJ...753...12K}. If necessary, we applied some arbitrary abundance adjustment to improve the model until the line ratios between the model and observation agree within at least 15\%. In most cases, the arbitrary adjustments were within the measurement and ICF uncertainties. The line strength comparison between the final model and the observation is presented in Appendix \ref{app:line_flux}. The abundance of the final model is described in Sect. \ref{result:nebular}. We employed the derived \teff~to minimise the most likely distance using $L_*$, $t_\mathrm{kin}$, and $t_\mathrm{tracks}$.

\subsection{$L-t_{kin}$ distance} \label{sec:ltkin_distance}

For a given \teff~from the photoionisation model, we simulated 4000 distance points between the minimum and maximum distance in the literature. The distance range was defined purely on the basis of computational efficiency, with no implications on the physical constraints. For each point, we calculated the $L_*$ and $t_\mathrm{kin}$ values. To make the comparison with the theoretical ages, we need a better resolution for the $t_\mathrm{tracks}$ as a function of initial mass, because the current spacing of the tracks by \citet{2016A&A...588A..25M} is scarce. To achieve this, we interpolated their stellar tracks; we generated 2000 steps between the mass range of $1-3 \:M_\odot$. The post-AGB model grid was computed at similar points in the HR-diagram, which allows a simple interpolation \citep{2016A&A...588A..25M}. We linearly interpolated the theoretical log \teff, log $L_*$, and $t_\mathrm{tracks}$ as a function of mass. Then, we assessed each distance point with $\Delta t = |t_\mathrm{kin} - t_\mathrm{tracks}|$. The distance associated with the minimum $\Delta t$ was then adopted as the $L-t_{kin}$ distance. The minimum was determined by fitting an asymmetric Gaussian function to the $\Delta t$ values as a function of distance. To estimate the uncertainty, we repeated the calculation of $L_*$ using the $L-t_{kin}$ distance. In the first iteration, the $L_*$ uncertainty was purely propagated from the $L_{\mathrm{H}\beta}$. Then, we identified the simulated distance points, which correspond to the value of $\pm dL_*$. We adopted the associated distance points as our $L-t_{kin}$ distance uncertainties.

The uncertainty of $\Delta t$ was negligible compared to the one stemming from $L_{\mathrm{H}\beta}$. As described in the previous subsection, the main uncertainty of $t_\mathrm{kin}$ was the determination of $v_\mathrm{exp}$, which typically need a factor of 1.25 correction \citep{2013A&A...558A..78J}. On the other hand, the uncertainty of the $t_\mathrm{tracks}$ was difficult to determine; the theoretical age is known to be a function of metallicity \citep{2016A&A...588A..25M} and we have selected the one closest to the solar metallicity. If we assumed an extreme of 50\% uncertainty for $t_\mathrm{kin}$, it would account for $<3- 10\%$ of the total $L-t_{kin}$ uncertainties.

Finally, we repeated the calculation of $L_*$ and $t_\mathrm{kin}$ with the $L-t_{kin}$ distance and its uncertainty; we adopted them as our fiducial results. Furthermore, we also refined our photoionisation models with the new $L_*$ value to obtain the fiducial \teff. The stellar parameters derived with the $L-t_{kin}$ distance, compared with other methods, is shown in Fig. \ref{fig:distance_tkin}. The parameter value comparison can be found in Table \ref{tab:distances}. For NGC\,6572 and NGC\,6884, our distance values are in agreement with the Gaia distances of CW21, with a slightly larger errors. The distance of M\,1-71 is larger than the available literature values, but more consistent with the theoretical expectations. On this basis, we adopt our $L_*-t_\mathrm{kin}$ values as our distances.

\section{Absolute magnitude $M_{5007}$} \label{sec:m5007}

To derive the absolute magnitude in \oiii, $M_{5007}$, firstly we need to calculate the apparent magnitude $m_{5007}$, which is defined by \citet{1989ApJ...339...39J} as 

\begin{equation}
    m_{5007} = -2.5 \: \mathrm{log} \: F_{5007} - 13.74,
\end{equation}

\noindent
where the $F_{5007}$ has the unit of erg~cm$^{-2}$~s$^{-1}$. The flux and its uncertainty were measured using our data in the same way as for the nebular diagnostic. Due to the limited FOV and some spaxel-masking in our observations, we lost some flux. While the flux loss was less relevant for the abundance measurement, as it relied on the line ratios, the flux loss needed to be quantified in this analysis. To estimate the flux loss, we simulated our FOV with the help of HST images of the PNe, which is described in detail in Appendix \ref{app:mosaic}. For NGC\,6572 and NGC\,6884, we estimated a loss of $\sim 1$ \% and $\sim 17$ \%, respectively. For M\,1-71, there is no HST image available and therefore we could not estimate the loss in the same manner. M\,1-71 was observed in the same night as NGC\,6884, right after it. Both PNe have a similar centre coordinate within the FOV. Although the angular size of M\,1-71 is smaller than the one of NGC\,6884, it was observed at a worse seeing condition; 1\farcs9 FWHM instead of 1\farcs5 FWHM during the observation of NGC\,6884. When we assumed the correction factor of NGC\,6884 for M\,1-71 and compared the corrected flux of M\,1-71 to literature values, they were in agreement. Therefore we adopted the same correction factor for M\,1-71, as for NGC\,6884. A comparison between our $F_{5007}$ and the literature is shown in Table \ref{tab:f5007}. Afterward, we calculated $m_{5007}$ for all the three nebulae and converted them into $M_{5007}$ using the distance modulus. 

\begin{table}
\caption{$F_{5007}$ of our PNe with the literature comparison.}
\label{tab:f5007}
\centering
\begin{tabular}{c l}
\hline
\hline
PN & log $F_{5007}$ [erg~cm$^{-2}$~s$^{-1}$]\\ 
\hline
\multirow{3}{*}{NGC\,6572} & $-8.68$ (C23)\\
          & $-8.68$ (Observed)\\
          & $-8.67$ (Flux loss corrected)\\
          
\hline
\multirow{4}{*}{NGC\,6884} & $-9.78$ (H97)\\
          & $-9.90$ (C23)\\
          & $-10.10$ (Observed)\\
          & $-10.01$ (Flux loss corrected)\\

\hline
\multirow{4}{*}{M\,1-71} & $-11.00$ (W05)\\
          & $-10.97$ (C23)\\
          & $-11.09$ (Observed)\\
          & $-11.01$ (Flux loss corrected)\\
\hline

\end{tabular}
\tablefoot{H97: \citet{1997ApJS..108..503H}, W05: \citet{2005A&A...436..967W}, C23: \citet{Chornay_Walton_Jones_Boffin_2023}}
\end{table}

In the context of the PNLF, we should correct for the foreground extinction but retain the circum-nebular extinction \citep{2025ApJ...983..129J}. This means that we cannot simply apply the correction value from our Balmer decrement; this value consists of the circum-nebular extinction and the Galactic foreground extinction. To measure the latter, we used \textsc{g-tomo}\footnote{https://github.com/explore-platform/g-tomo} that employed dust extinction cubes from \citet{2022A&A...661A.147L, 2022A&A...664A.174V}. We used the extinction map with the volume of 6 $\times$ 6 $\times$ 0.8 kpc. The Galactic foreground extinction was determined based on the positional coordinates of the PNe and the adopted distances. Once the Galactic foreground extinction is known, we can also calculate the circum-nebular extinction with $c(\mathrm{H}\beta)_\mathrm{neb} = c(\mathrm{H}\beta)_\mathrm{total} - c(\mathrm{H}\beta)_\mathrm{Galactic}$. 

From our foreground extinction corrected $M_{5007}$ values, we will confirm whether our PNe are indeed part of the PNLF cutoff. We will also put a constrain on the circum-nebular extinction and see whether our derived central star masses are in agreement with the relation between final mass and circum-nebular extinction relation at the PNLF cutoff from \citet{2025ApJ...983..129J}. We can also check if the conversion efficiency of the central star $L_*$ into $L_{5007}$ of our PNe are in an agreement with what is observed in the nearby galaxies. This will be discussed in Sect. \ref{sec:cspn_mass}. 

\section{Results} \label{sec:result}

\subsection{Nebular properties}
\label{result:nebular}

We present $c(\mathrm{H\beta)}$, $T_e$, and $N_e$ values of our PNe in Table \ref{tab:nebular_diagnostics}. $T_e$[\ion{N}{ii}] and $T_e$[\ion{O}{iii}] of our sample have a narrow range of $\sim$12500--15300 K and $\sim$11350 --11700 K, respectively. The $N_e$ of each diagnostic lines for our PNe also have similar values within the uncertainties. The measured $c(\mathrm{H\beta)}$ of M\,1-71 has the largest extinction among our sample. 

The ionic and elemental abundances are presented in Table \ref{tab:ionic_abund} and Table \ref{tab:elemental_abund}, respectively. The elemental abundances that we adopted for our best \textsc{Cloudy} models and the literature comparison are also shown in Table \ref{tab:elemental_abund}. We found that the calculation of the total nitrogen abundance really depends on the adopted ICF scheme. In this case, our \textsc{Cloudy} models are in a better agreement with the KB94 scheme. The abundance comparison between each PN does not show any significant difference. Overall, our PNe exhibit the abundances of typical MW PNe \citep{2022PASP..134b2001K, 2025arXiv251014149E}. 

\begin{table}
\caption{Values of $c(\mathrm{H}\beta)$, $T_e$, and $N_e$ for our PN sample. The measured $c(\mathrm{H}\beta)_\mathrm{total}$ values consist of the foreground and circum-nebular extinction.}
\label{tab:nebular_diagnostics}
\centering
\begin{tabular}{c c c c}
\hline
\hline
Parameter & NGC\,6572 & NGC\,6884 & M\,1-71 \\
\hline
\hline
$c(\mathrm{H}\beta)_\mathrm{total}$ & 0.46$_{-0.15}^{+0.14}$  & 0.76$_{-0.11}^{+0.12}$ & 2.13$_{-0.12}^{+0.12}$\\
\hline

$T_e$[\ion{N}{ii}] [K]& 12950$_{-310}^{+290}$ & 12490$_{-250}^{+250}$ & 15300$_{-370}^{+470}$ \\
$T_e$[\ion{O}{iii}] [K]& 11400$_{-250}^{+290}$ & 11680$_{-220}^{+210}$ & 11350$_{-280}^{+310}$\\
\hline

log $N_e$[\ion{S}{ii}] cm$^{-2}$& 4.22$_{-0.33}^{+0.46}$ & 3.90$_{-0.27}^{+0.36}$ & 4.11$_{-0.26}^{+0.30}$\\
log $N_e$[\ion{Cl}{iii}] cm$^{-2}$& 3.95$_{-0.10}^{+0.12}$ & 3.63$_{-0.11}^{+0.10}$ & --\\
log $N_e$[\ion{Ar}{iv}] cm$^{-2}$& 3.75$_{-0.02}^{+0.02}$ & 3.74$_{-0.03}^{+0.02}$ & 3.81$_{-0.15}^{+0.15}$ \\
\hline
\end{tabular}
\end{table}

\begin{table}
\caption{Ionic abundances. He$^{+}$ and He$^{++}$ are presented as $N(X^i)/N(H^+)\times10$. S$^{+}$, C$^{++}$, Ar$^{++}$, and Ar$^{+++}$ are presented as $N(X^i)/N(H^+)\times10^{7}$. The rest is presented as $N(X^i)/N(H^+)\times10^{5}$.}
\label{tab:ionic_abund}
\centering
\begin{tabular}{l c c c c}
\hline
\hline
Ion & $\lambda\:[\AA]$ & NGC\,6572 & NGC\,6884 & M\,1-71 \\
\hline
\hline
He$^{+}$ & 5876 & 1.04$_{-0.11}^{+0.13}$ & 0.98$_{-0.07}^{+0.07}$ & 1.15$_{-0.08}^{+0.09}$ \\
        & 6678 & 1.01$_{-0.08}^{+0.09}$  & 0.98$_{-0.08}^{+0.10}$ & 1.07$_{-0.10}^{+0.11}$\\
        & adopted & 1.02$_{-0.09}^{+0.11}$ & 0.95$_{-0.08}^{+0.09}$ & 1.12$_{-0.11}^{+0.10}$\\
\hline

He$^{++}$& 4686 & 0.008$_{-0.0004}^{+0.0004}$ & 0.20$_{-0.01}^{+0.01}$ & --\\
\hline

N$^{+}$ & 5755 &  1.02$_{-0.08}^{+0.09}$  & 0.51$_{-0.04}^{+0.04}$ & 0.73$_{-0.05}^{+0.06}$\\
        & 6548 &  0.99$_{-0.10}^{+0.12}$  & 0.53$_{-0.04}^{+0.05}$ & 0.77$_{-0.08}^{+0.08}$\\
        & 6584 &  1.01$_{-0.11}^{+0.14}$  & 0.52$_{-0.05}^{+0.05}$ & 0.77$_{-0.07}^{+0.07}$\\
        & adopted & 1.01$_{-0.10}^{+0.11}$ & 0.52$_{-0.04}^{+0.05}$ & 0.76$_{-0.06}^{+0.07}$\\
\hline

O$^{0}$ & 6300 &  0.69$_{-0.07}^{+0.07}$ & 0.42$_{-0.03}^{+0.04}$ & 0.44$_{-0.04}^{+0.04}$\\
        & 6363 &  0.73$_{-0.08}^{+0.08}$  & 0.46$_{-0.04}^{+0.04}$ & 0.48$_{-0.04}^{+0.05}$\\
        & adopted & 0.70$_{-0.08}^{+0.09}$ & 0.44$_{-0.04}^{+0.04}$ & 0.46$_{-0.05}^{+0.05}$\\
\hline

O$^{+}$ & 3727 &  2.31$_{-0.28}^{+0.30}$ & 0.92$_{-0.09}^{+0.10}$ & 1.26$_{-0.12}^{+0.14}$\\
        & 7320 &  --  & 0.67$_{-0.07}^{+0.08}$ & 0.86$_{-0.09}^{+0.10}$\\
        & 7330 &  --  & 0.71$_{-0.07}^{+0.09}$ & 0.89$_{-0.10}^{+0.12}$\\
        & adopted & 2.31$_{-0.28}^{+0.30}$ & 0.75$_{-0.11}^{+0.24}$ & 0.87$_{-0.10}^{+0.12}$\\
\hline

O$^{++}$ & 5007 &  25.99$_{-1.39}^{+1.48}$  & 28.21$_{-1.37}^{+1.28}$ & 26.20$_{-1.54}^{+1.48}$\\
        
\hline

Ne$^{++}$ & 3869 & 7.79$_{-0.86}^{+0.92}$ & 8.86$_{-0.75}^{+0.84}$ & 12.96$_{-1.22}^{+1.32}$\\
\hline

S$^{+}$ & 6716 &  1.76$_{-0.19}^{+0.24}$  & 1.69$_{-0.16}^{+0.17}$ & 1.21$_{-0.10}^{+0.12}$\\
        & 6731 &  1.95$_{-0.23}^{+0.25}$  & 1.71$_{-0.17}^{+0.16}$ & 1.24$_{-0.11}^{+0.13}$\\
        & adopted & 1.86$_{-0.23}^{+0.25}$ & 1.70$_{-0.16}^{+0.17}$ & 1.22$_{-0.11}^{+0.13}$\\
\hline

S$^{++}$ & 6312 &  0.18$_{-0.02}^{+0.02}$  & 0.22$_{-0.02}^{+0.02}$  & 0.14$_{-0.01}^{+0.01}$\\
\hline

Cl$^{++}$ & 5518 & 0.48$_{-0.03}^{+0.04}$ & 0.59$_{-0.04}^{+0.03}$ & --\\
        & 5538 & 0.60$_{-0.04}^{+0.04}$ & 0.67$_{-0.04}^{+0.04}$ & 0.54$_{-0.03}^{+0.04}$ \\
        & adopted & 0.53$_{-0.08}^{+0.09}$ & 0.62$_{-0.06}^{+0.07}$ & 0.54$_{-0.03}^{+0.04}$\\
\hline

Ar$^{++}$ & 7135 & -- & 9.45$_{-1.00}^{+0.98}$ & 10.71$_{-1.15}^{+1.40}$\\
\hline

Ar$^{+++}$ & 4712 & 5.11$_{-0.25}^{+0.28}$ & 11.46$_{-0.58}^{+0.61}$ & 4.59$_{-0.23}^{+0.23}$\\
        & 4740 & 5.15$_{-0.28}^{+0.26}$ & 11.45$_{-0.58}^{+0.56}$ & 4.89$_{-0.24}^{+0.26}$\\
        & adopted & 5.12$_{-0.26}^{+0.27}$ & 11.46$_{-0.58}^{+0.58}$ & 4.73$_{-0.27}^{+0.31}$\\
\hline

\end{tabular}
\end{table}

\begin{table*}
\caption{Elemental abundances as $12 + \mathrm{log_{10}}(X/H)$ for our observation and \textsc{Cloudy} models. The comparison to literature values is also presented.}
\label{tab:elemental_abund}
\centering
\begin{tabular}{l c c c c c |c c c c |c c c c}
\hline
\hline
\multirow{2}{*}{Element} & \multirow{2}{*}{Ref$_\mathrm{ICF}$} & \multicolumn{4}{c}{NGC\,6572} & \multicolumn{4}{c}{NGC\,6884} & \multicolumn{3}{c}{M\,1-71} \\

& & ICF & This work & H94 & B23 & ICF & This work & H97 & L04 & ICF & This work & W05\\
\hline
\hline

\multirow{3}{*}{Helium} & \multirow{2}{*}{BSJ} & \multirow{2}{*}{1.00} & \multirow{2}{*}{11.02$_{-0.04}^{+0.04}$} & \multirow{3}{*}{11.04} & \multirow{3}{*}{10.93} & \multirow{2}{*}{1.00} &\multirow{2}{*}{11.07$_{-0.03}^{+0.03}$} & \multirow{3}{*}{10.99} & \multirow{3}{*}{11.00} & \multirow{2}{*}{1.00} & \multirow{2}{*}{11.05$_{-0.04}^{+0.04}$}& \multirow{3}{*}{11.07}\\

                     &&&&&&&&&&&\\
                     & \multicolumn{2}{c}{\textsc{Cloudy}} & 11.02 &&&& 11.07 &&&& 11.05\\
\hline

\multirow{3}{*}{Nitrogen} & KB94 & 12.32 & 8.09$_{-0.08}^{+0.07}$ & \multirow{3}{*}{7.79} & \multirow{3}{*}{8.16} & 38.80 & 8.31$_{-0.08}^{+0.10}$ & \multirow{3}{*}{8.47} & \multirow{3}{*}{8.15} & 28.57 & 8.33$_{-0.09}^{+0.13}$ & \multirow{3}{*}{--}\\
                     
                     & DI14 & 1.46 & 7.17$_{-0.19}^{+0.88}$ &&& --&-- &&& 4.15 & 7.50$_{-0.05}^{+0.06}$ \\
                     & \multicolumn{2}{c}{\textsc{Cloudy}} & 7.97 &&&& 8.25 &&&& 8.15 \\
\hline

\multirow{3}{*}{Oxygen} & KB94 & 1.00 & 8.46$_{-0.03}^{+0.02}$ & \multirow{3}{*}{8.54} & \multirow{3}{*}{8.54} & 1.13 & 8.52$_{-0.02}^{+0.02}$ & \multirow{3}{*}{8.95} & \multirow{3}{*}{8.61} & 1.00 & 8.43$_{-0.03}^{+0.02}$ & \multirow{3}{*}{8.70}\\
                     & DI14 & 1.00 & 8.46$_{-0.03}^{+0.02}$ &&&1.11 & 8.51$_{-0.02}^{+0.02}$ &&& 1.00 & 8.43$_{-0.03}^{+0.02}$\\
                     & \multicolumn{2}{c}{\textsc{Cloudy}} & 8.47 &&&& 8.53 &&&& 8.43  \\
\hline

\multirow{3}{*}{Neon} & KB94 & 1.09 & 7.93$_{-0.07}^{+0.06}$ & \multirow{3}{*}{7.76} & \multirow{3}{*}{8.05} & 1.02 & 7.99$_{-0.05}^{+0.04}$ & \multirow{3}{*}{8.43} & \multirow{3}{*}{8.16} & 1.04 & 8.13$_{-0.06}^{+0.05}$ & \multirow{3}{*}{8.18}\\
                     & DI14 & 2.27 & 8.25$_{-0.06}^{+0.05}$ &&& 0.97 & 7.94$_{-0.09}^{+0.06}$ &&& 1.37 & 8.25$_{-0.07}^{+0.05}$\\
                     & \multicolumn{2}{c}{\textsc{Cloudy}} & 7.97 &&&& 8.05 &&&& 8.20\\
\hline
                     
\multirow{3}{*}{Sulphur} & KB94 & 1.65 & 6.53$_{-0.04}^{+0.04}$ & \multirow{3}{*}{6.33} & \multirow{3}{*}{6.60} & 2.37 & 6.75$_{-0.03}^{+0.03}$& \multirow{3}{*}{6.81}& \multirow{3}{*}{6.80} & 2.14 & 6.49$_{-0.10}^{+0.09}$ & \multirow{3}{*}{--}\\
                     & DI14 & 1.00 & 6.31$_{-0.04}^{+0.04}$ &&& 1.00 & 6.60$_{-0.02}^{+0.02}$ &&& 2.04 & 6.51$_{-0.04}^{+0.03}$\\
                     & \multicolumn{2}{c}{\textsc{Cloudy}} & 6.35 &&&& 6.50 &&&& 6.40\\
\hline

\multirow{3}{*}{Chlorine} & KB94 & - & - & \multirow{3}{*}{4.83} & \multirow{3}{*}{5.05} & -- & -- &\multirow{3}{*}{5.24}&\multirow{3}{*}{5.20}& \multirow{3}{*}{2.91}& \multirow{3}{*}{5.20$_{-0.05}^{+0.04}$}& \multirow{3}{*}{--}\\

                     & DI14 & 2.09 & 5.05$_{-0.09}^{+0.07}$ &&& 2.91 & 5.26$_{-0.06}^{+0.05}$ & &  \\
                     & \multicolumn{2}{c}{\textsc{Cloudy}} & 4.95 &&&& 5.10 &&&& 5.10 \\
\hline

\multirow{3}{*}{Argon} & KB94 & 1.09 & 5.75$_{-0.09}^{+0.07}$ & \multirow{3}{*}{6.28} & \multirow{3}{*}{6.34} & 1.87 & 6.25$_{-0.03}^{+0.02}$ & \multirow{3}{*}{6.42} & \multirow{3}{*}{6.26} & 1.87 & 6.30$_{-0.03}^{+0.03}$& \multirow{3}{*}{--} \\
                     & DI14 & - & - &&& 1.66 & 6.20$_{-0.25}^{+0.14}$ &&& 1.63 & 6.24$_{-0.25}^{+0.14}$ \\
                     & \multicolumn{2}{c}{\textsc{Cloudy}} & 6.35 &&&& 6.25 &&&& 6.22\\
\hline

\end{tabular}
\end{table*}

\subsection{Central star properties} \label{sec:cspn_param}

The $L_*$ and \teff~of our CSPNe from the \textsc{Cloudy} models are presented in Table \ref{tab:cspn_lum_teff}. For NGC\,6572 and NGC\,6884, we compared the values with previous studies by \citet[][BD23]{2023MNRAS.524.1547B} and \citet[][H97]{1997ApJS..108..503H}, respectively; no model is available in the literature for M\,1-71. 

As mentioned before, our models have more simplistic assumptions compared to those mentioned in the literature. The photoionisation model of NGC\,6572 by BD23 assumed a bipolar nebular shell, which was derived from the morph-kinematic model of the nebula. They adopted a distance of 1.85 kpc, which was $\sim100$ pc further than our distance. Nevertheless, our $L_*$ and \teff~of NGC\,6572 are in agreement with BD23.

For NGC\,6884, however, we derived a higher $L_*$ and \teff~compared to H97. The model of NGC\,6884 by H97 was constructed using a composite-shell geometry, with the assumed distance of 2 kpc, which was almost 3 kpc shorter than our adopted value. To fully explain the discrepancy of the modelling aspect, one needed to repeat the analysis of exact model assumption with our abundance and distance values; this is beyond the scope of the current study. Since our focus is to compare these Galactic PNe to their extragalactic counterparts, that are typically modelled using the basic spherical assumption, our current analysis should suffice. 

The initial mass of our CSPNe are shown in Table \ref{tab:cspn_lum_teff}. The values were derived through the interpolation of the stellar tracks, with respect to the position of each object in the HR-diagram. NGC\,6572 and NGC\,6884 have a similar progenitor mass within their uncertainties, while M\,1-71 has a slightly larger initial mass. For NGC\,6572, our initial mass is in agreement with the initial mass derived by BD23, who also used the tracks of \citet{2016A&A...588A..25M}. On the other hand, our initial mass value of NGC\,6884 is larger than H97, which can most likely be attributed to the discrepant $L_*$, \teff, and their stellar tracks; they adopted the post-AGB models from \citet{1994ApJS...92..125V}. 

\begin{table*}
\caption{Central star parameters.}
\label{tab:cspn_lum_teff}
\centering
\begin{tabular}{c c c c c c}
\hline
\hline
\multirow{2}{*}{Parameter} & \multicolumn{2}{c}{NGC\,6572} & \multicolumn{2}{c}{NGC\,6884} & M\,1-71 \\
                        & This work & BD23 & This work & H97 & This work \\
\hline
\hline
log $T_\mathrm{eff}$ [K] & 4.85 & 4.83 & 5.11 & 5.04 & 4.90\\
\hline
log $L$ $[L_\odot]$& 3.76$_{-0.19}^{+0.22}$ & 3.76 & 3.62$_{-0.21}^{+0.18}$ & 3.34 & 3.81$_{-0.19}^{+0.19}$\\
\hline
$M_\mathrm{init}$ $[M_\odot]$& 1.36$_{-0.25}^{+1.12}$ & 1.25 & 1.37$_{-0.18}^{+0.72}$ & 1.00 & 1.73$_{-0.55}^{+0.93}$\\
\hline
\end{tabular}
\end{table*}

\subsection{$M_{5007}$ values} \label{sec:M5007}

As explained in Sect. \ref{sec:m5007}, our Balmer decrement measured the sum of the Galactic foreground extinction and the circum-nebular extinction. The Galactic foreground extinction derived using \textsc{g-tomo}, the circum-nebular extinction, and the $M_{5007}$ values for our sample can be seen in Table \ref{tab:m5007}. We found that the Galactic foreground extinction has a significant contribution to the total measured extinction for all PNe. The uncertainty of the $c(\mathrm{H}\beta)_\mathrm{neb}$ is dominated by our measurement uncertainties in the Balmer decrement. 

For distance determination of external galaxies, the typical assumption of the PNLF cutoff is $M^* = -4.53 \pm 0.06$ \citep{2012Ap&SS.341..151C, 2022FrASS...9.6326C}. Based on our derived $M_{5007}$, our PNe should be located near the top $\sim1$ mag of the PNLF; this magnitude region is typically observed in extragalactic PNLF up to a distance of $\sim30$ Mpc \citep{2021A&A...653A.167S, 2021ApJ...916...21R, 2024ApJS..271...40J, 2025A&A...704A.303S}. NGC\,6572 has the brightest $M_{5007}$ among the three objects, followed by NGC\,6884 and then M\,1-71. To see whether our sample shares the characteristics of PNe in extragalactic studies, we discuss them in the next section. Moreover, unlike their extragalactic counterparts, we can also discuss the most likely central star type of the PNe to infer their evolution. 

\begin{table}
\caption{Extinction components and $M_{5007}$ of our PNe.}
\label{tab:m5007}
\centering
\begin{tabular}{c c c c}
\hline
\hline
PN & $c(\mathrm{H}\beta)_\mathrm{Galactic}$ & $c(\mathrm{H}\beta)_\mathrm{nebula}$ & $M_{5007}$ \\
\hline
\hline
NGC\,6572 & $0.34\pm0.01$ & $0.12_{-0.15}^{+0.14}$ & $-4.18_{-0.41}^{+0.34}$\\
\hline
NGC\,6884 & $0.67\pm0.01$ & $0.09_{-0.11}^{+0.12}$ & $-4.01_{-0.38}^{+0.41}$\\
\hline
M\,1-71 & $1.75\pm0.15$ & $0.38_{-0.13}^{+0.18}$ & $-3.60_{-0.69}^{+0.48}$\\
\hline
\end{tabular}
\end{table}

\begin{figure*}[]
   \centering
   \includegraphics[width=0.85\linewidth]{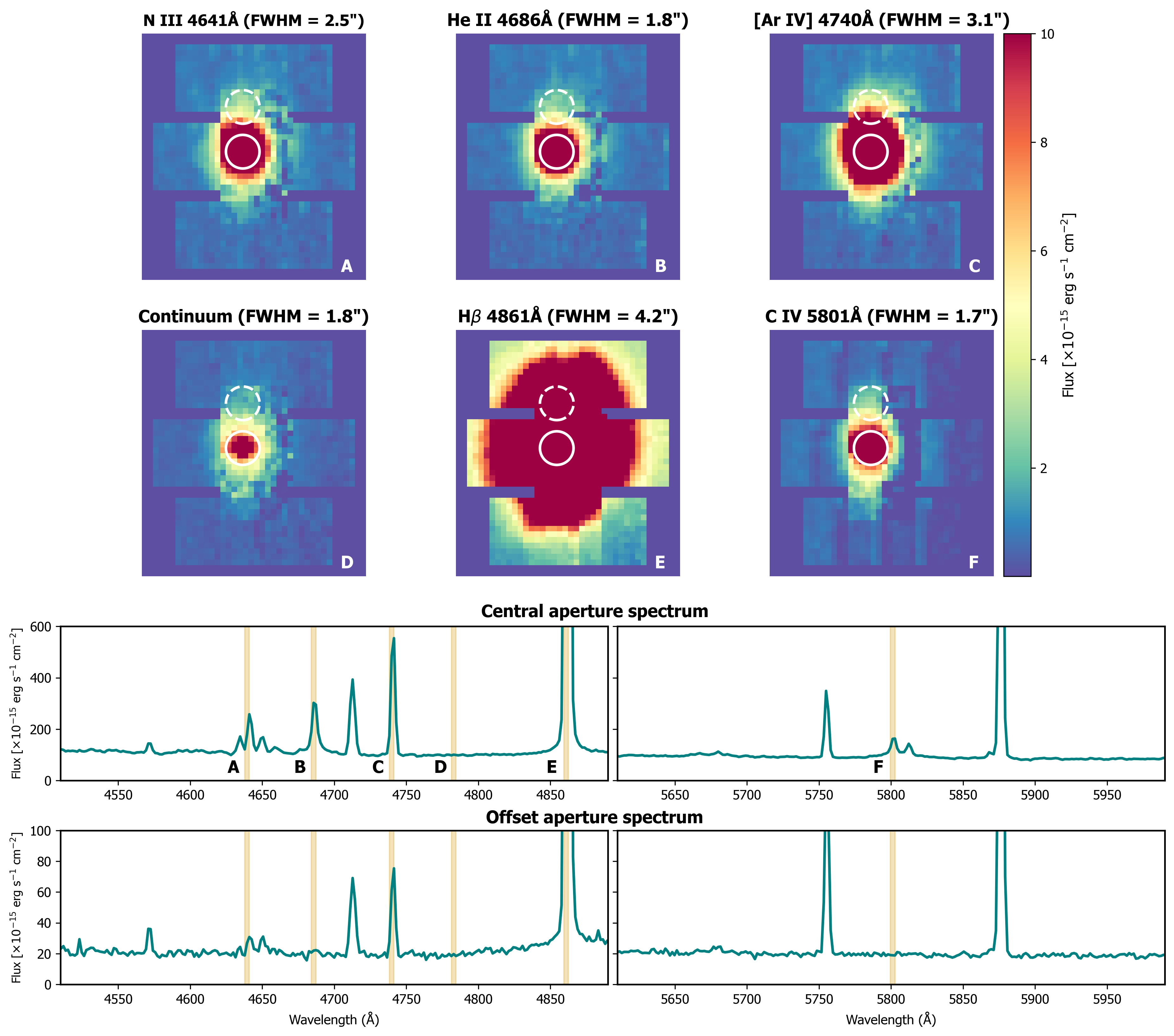}
   \caption{Emission line maps and partial spectra of NGC\,6572. The two upper panels show six emission line maps with their image FWHMs in brackets: three \textit{wels} line maps, an \hb~map, a high ionisation zone map of [\ion{Ar}{iv}]$\lambda4740$, and a continuum map. The solid circle marks the aperture centred on the CSPN position; the spectra is shown in the third panel. The dashed circle marks the aperture off-setted from the CSPN position; the spectra is shown in the fourth panel. Shaded regions in the middle and lower panel indicate the position of the emission line maps in the spectra. }
   \label{fig:wels_ngc6572}
\end{figure*}

\begin{figure*}[]
   \centering
   \includegraphics[width=0.85\linewidth]{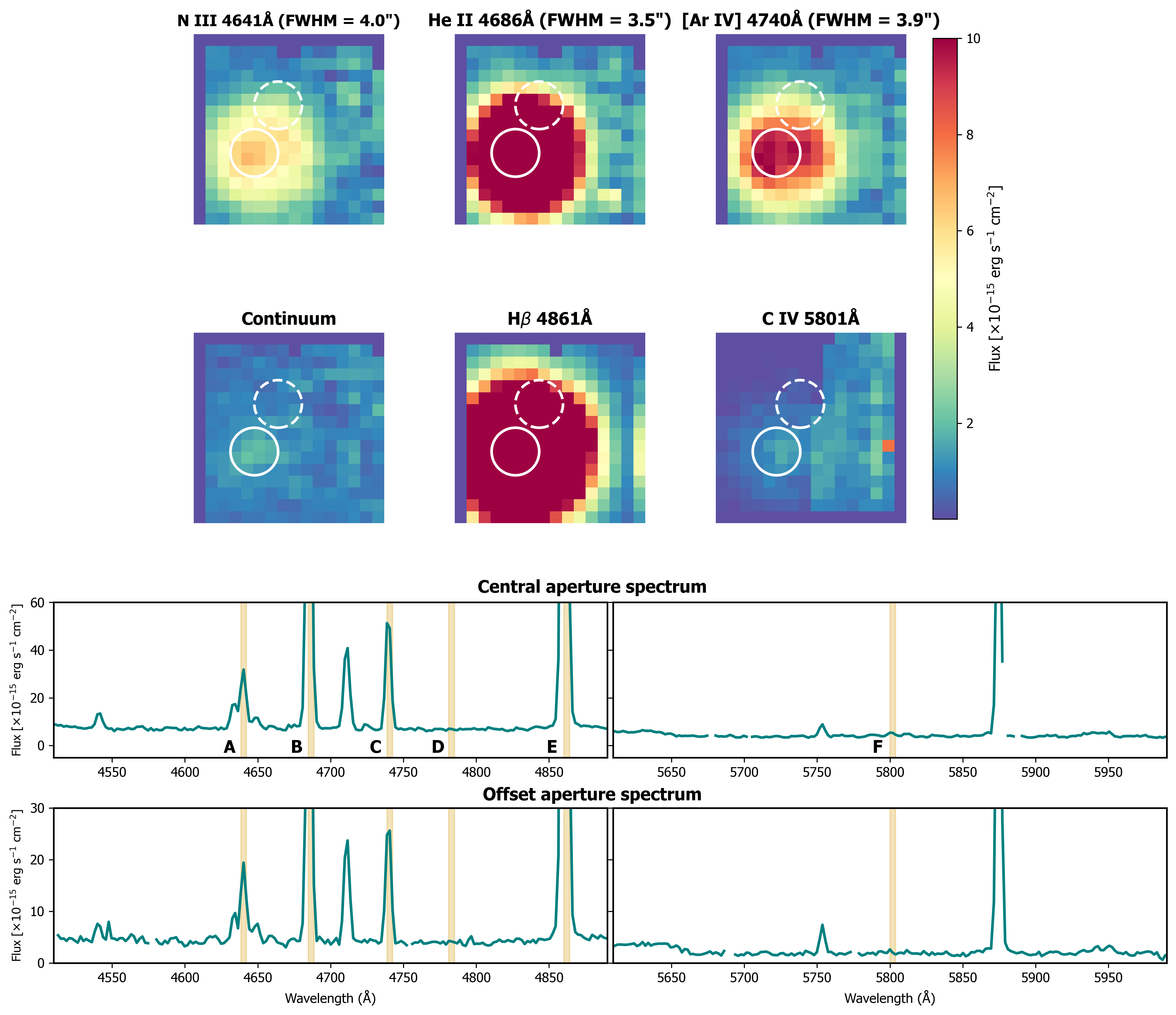}
   \caption{Same as Fig. \ref{fig:wels_ngc6572} for NGC\,6884. The image FWHMs of the \ion{C}{iv}$\lambda5801$ and the continuum images were not derived due to low signal. The FWHM of the \hb~image was not calculated due to image truncation.}
   \label{fig:wels_ngc6884}
\end{figure*}

\begin{figure*}[]
   \centering
   \includegraphics[width=0.85\linewidth]{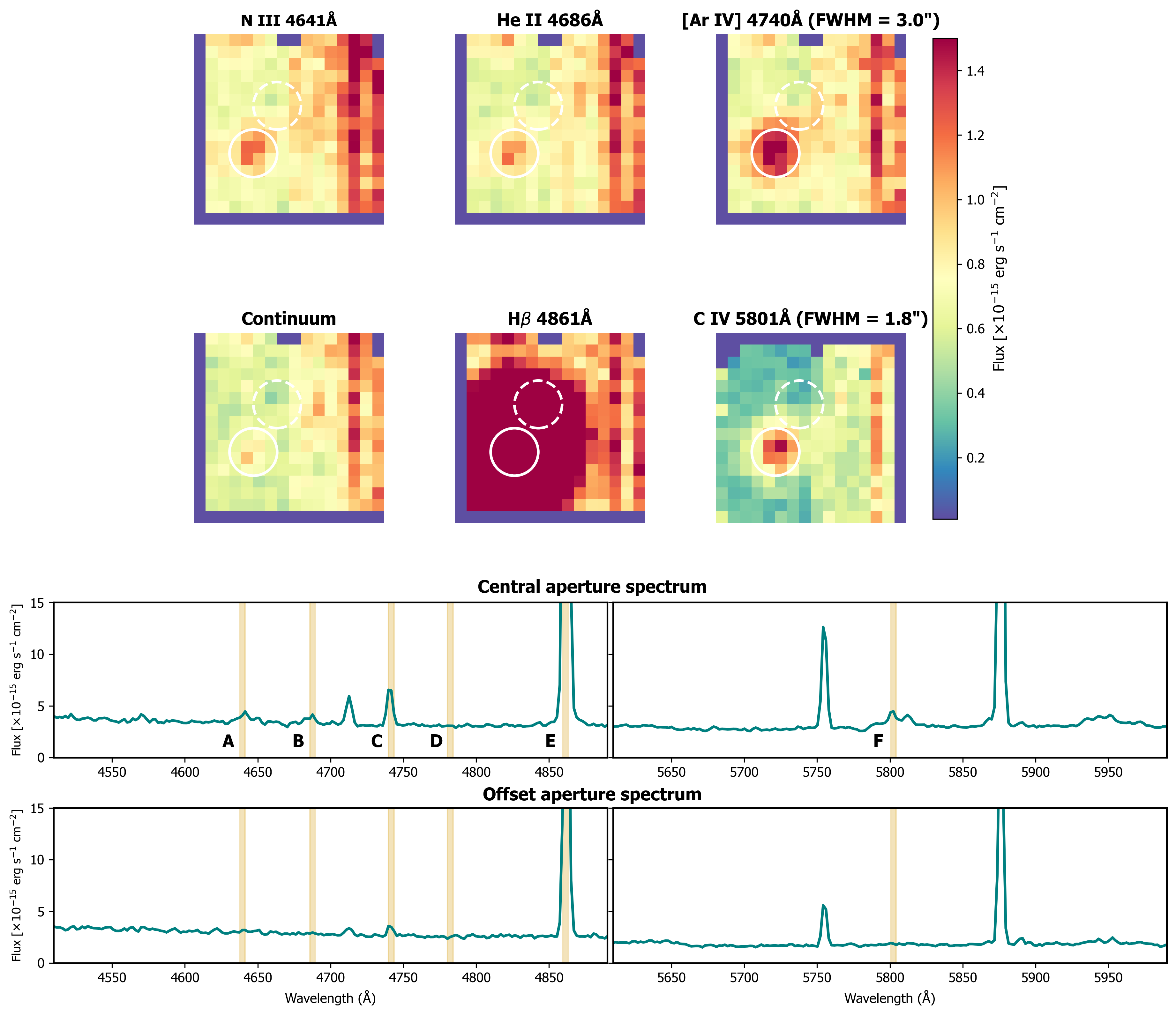}
   \caption{Same as Fig. \ref{fig:wels_ngc6572} for M\,1-71. The image FWHMs of the \ion{N}{iii}$\lambda4634-41$ image, the \heii~image, and the continuum image were not determined due to low signal. Similar to NGC\,6884, the the \hb~image was also truncated and therefore the FWHM value was not calculated.}
   \label{fig:wels_m171}
\end{figure*}

\section{Discussions} \label{sec:discuss}

\subsection{Central star classification} \label{sec:cs_class}

In earlier investigations, the central stars of NGC\,6572 and M\,1-71 were classified as weak-emission line star \citep[\textit{wels},][]{1998A&A...329L...9P}. The \textit{wels} is typically characterised by the detection of narrow emission lines of \ion{N}{iii}$\lambda4634-41$, \ion{C}{iii}$\lambda4647-50$, \ion{C}{iv}$\lambda4658$, \heii, and \ion{C}{iv}$\lambda5801-12$ \citep{1993A&AS..102..595T, 2003AJ....126..887M}. These features are consistent with those from Wolf-Rayet central stars ([WR]), whose lines, however, are prominently stronger and broader. For NGC\,6884, \citet{1997ApJS..108..503H} assigned a central star type [WN], i.e. a nitrogen-rich Wolf-Rayet star, but no further details were explained. We speculate that this may be due to the detection of \ion{N}{iii}$\lambda4634-41$ emission line, which is typically observed in classical WN stars \citep[e.g.,][]{2001NewAR..45..135V, 2017A&A...603A.130M}. Using our data, we can identify those lines and spatially determine their region of emission. 

The spectral and spatial inspection for the emission lines is shown in Fig. \ref{fig:wels_ngc6572} (NGC\,6572), Fig. \ref{fig:wels_ngc6884} (NGC\,6884), and Fig. \ref{fig:wels_m171} (M\,1-71). We inspected the emission lines of \ion{N}{iii}$\lambda4634-41$, \heii, and \ion{C}{iv}$\lambda5801-12$. As a comparison, we show a narrow continuum image to indicate the position of the central star.  The \hb~and [\ion{Ar}{iv}]$\lambda4740$ maps illustrate the extension of the PNe. Each narrowband image has a width of 6 \AA, centred at the line peak. We did not generate a continuum image with a larger wavelength range to minimise the contribution from the nebular continuum. To explain the origin of these lines, we analysed two full width at half maximum (FWHM) measurements: PN image FWHMs, obtained by fitting Moffat profiles to selected images (Fig. \ref{fig:wels_ngc6572}, \ref{fig:wels_ngc6884}, and \ref{fig:wels_m171}), and PN emission line FWHMs, obtained by fitting Gaussian profiles. The measured line FWHMs for each PN are listed in Appendix \ref{app:line_flux}.

For NGC\,6572, the continuum image exhibited FWHM of 1.8\arcsec, which we assumed to be consistent with the point spread function (PSF) FWHM of the central star. We note that this value is larger than the measured seeing FWHM of the guiding camera of $\sim$1.5\arcsec at the end of the observation, which can be explained by the wavelength dependence of seeing and the different effective wavelengths of the continuum band near 4775 \AA, versus the V-band of the guiding camera. For the \ion{N}{iii}$\lambda4641$ line, we derived the image FWHM of 2.5\arcsec, indicating a more extended nature and originated from the nebula. The \heii~and \ion{C}{iv}$\lambda5801$ image FWHMs are 1.8\arcsec and 1.7\arcsec, respectively. This is smaller than the image FWHM of the high ionisation line [\ion{Ar}{iv}]$\lambda4740$, which has the value of 3.1\arcsec. Moreover, the \heii~and \ion{C}{iv}$\lambda5801$ have line FWHMs of 4.3 \AA~and 6.1 \AA, respectively. These values are broader than those of other lines, with a typical FWHM of $\sim 3.0$ \AA, as presented in Table \ref{tab:data_table_NGC6572}. Specifically for the \heii~line, we could not reproduce the intensity by considering only nebular contribution. To match the observed \heii~intensity, we needed to increase the temperature from log \teff$\:=4.85$ to log \teff$\:=4.99$. However, using the same abundances, the agreement of several important line intensities is no longer achieved. For example, the \oiii~intensity would be overestimated by $\sim25\%$; a minor abundance adjustment within the measurement uncertainties did not suffice to account the discrepancy. We speculate that the \heii~and the \ion{C}{iv}$\lambda5801$ lines originate from, or are predominantly contributed by, the central star.

The images of NGC\,6884 show a nebular origin of the \ion{N}{iii}$\lambda4641$ and \heii~lines, with image FWHMs of 4.0\arcsec and 3.5\arcsec, respectively. These are comparable to the image FWHM of the high ionisation zone traced by [\ion{Ar}{iv}]$\lambda4740$. Note that the seeing FWHM during the observation was $\sim$1.5\arcsec. Due to low signal, we did not fit the \ion{C}{iv}$\lambda5801$ and the continuum images. The \hb~image was truncated, so we did not calculate its FWHM. The emission line FWHMs of NGC\,6884 do not show any peculiarities; this can be referred in Table \ref{tab:data_table_NGC6884}.

We only fitted the \ion{C}{iv}$\lambda5801$ and the [\ion{Ar}{iv}]$\lambda4740$ images for M\,1-71. We derived a FWHM of 1.8\arcsec for the \ion{C}{iv}$\lambda5801$ image, which is comparable to the seeing FWHM of $\sim$1.9\arcsec. As a comparison, the [\ion{Ar}{iv}]$\lambda4740$ exhibits an image FWHM of 3.0\arcsec. For the other images, although we are able to see the features in the integrated spectrum of the central aperture, we were not able to perform any meaningful fit due to the low signal, or truncation in the case of the \hb~image. As presented in Table \ref{tab:data_table_M171}, the line FWHMs of M\,1-71 do not show any anomalies. We did not fit the line FWHM of \ion{C}{iv}$\lambda5801$, due to lower signal and spectral resolution compared to the NGC\,6572 data. Nevertheless, we speculate that, similar to NGC\,6572, the \ion{C}{iv}$\lambda5801$ emission likely originated from the central star.


Initially, \textit{wels} was suggested to be an evolutionary sequence of H-poor central star, bridging between the [WR] phase and the PG1159 phase, mostly due to the striking similarity with [WR] spectra \citep{1998A&A...329L...9P, 2003AJ....126..887M}. However, some authors \citep{2012IAUS..283..107M, 2014RMxAA..50..203K, 2015A&A...579A..86W} argued that the \textit{wels} classification is mostly caused by the insufficiency of low-resolution spectra. Specifically, \citet{2015A&A...579A..86W} found that most \textit{wels} classifications turned into H-rich O-type stars, when classified with high-resolution spectra; this included one of our PN, NGC\,6572. \citet{2016MNRAS.458.2694B} did a study of four \textit{wels} PNe and speculated that the \textit{wels} emission lines may emerge from a carbon- and nitrogen-rich gas pockets in the nebulae. This was confirmed with the spatial distribution of \textit{wels} features in the nebulae, which we also confirmed in the case of the \ion{N}{iii}$\lambda4641$ line in NGC\,6572 and NGC\,6884. They derived a progenitor mass range of $1.50 - 2.00 \:M_\odot$ for their sample, based on stellar tracks of \citet{1994ApJS...92..125V}. This is slightly larger to our progenitor mass range of $1.30 - 1.75 \:M_\odot$. However, if we compare their central star parameters to the stellar tracks of \citet{2016A&A...588A..25M}, they would have a similar progenitor mass, as shown in Fig. \ref{fig:HR_diagram_comparison}. Currently, we can not tell whether this is a general trend or simply a coincidence. 

Furthermore, based on the infrared study of the dust content of PNe, \citet{2024arXiv241211721M} suggested that most \textit{wels} PNe were formed directly from the AGB phase, without undergoing any late or very late thermal pulses; such a process is responsible for the formation of some of the known H-poor central stars \citep{1993PASP..105.1373I, 2019MNRAS.489.1054L}. All this information combined, it is very likely that NGC\,6884 and M\,1-71 also have H-rich central stars, but follow-up with high-resolution spectroscopy is necessary to confirm. Currently, there is more evidence that points to the H-rich nature of the central star in our sample than the opposite, and so the assumption of H-rich post-AGB tracks in our analysis is more justified. This supports the use of H-rich tracks in simulation studies of the PNLF \citep{2019ApJ...887...65V, 2025A&A...699A.371V} and the analysis of extragalactic PNe. 

\subsection{$M_{5007}$ evolutionary stage} \label{sec:oiii_evolution}

We determined the evolutionary stage of $M_{5007}$ using the excitation class by comparing it to the simulations in the literature. Based on 1-D hydrodynamical simulation, \citet{2007A&A...473..467S} predicted that $M_{5007}$ reaches its peak brightness at the excitation class (EC) of $\sim 4-6$; see their Fig. 13 and Fig. 14 for details. Then, as the nebulae evolve, the EC would continue to increase while $M_{5007}$ fades. The peak EC prediction is consistent with the observation of PNe in the Large Magellanic Cloud \citep[LMC,][]{1991ApJ...367..115D,1991ApJ...377..480D}, where the \oiii-luminous PNe have EC $\sim 5-6$. Recently, \citet{2022A&A...657A..71G} has shown that their sample of most \oiii-luminous PNe in M31 have EC values of $\sim4.5-6.3$. 

\begin{figure}[]
   \centering
   \includegraphics[width=\linewidth]{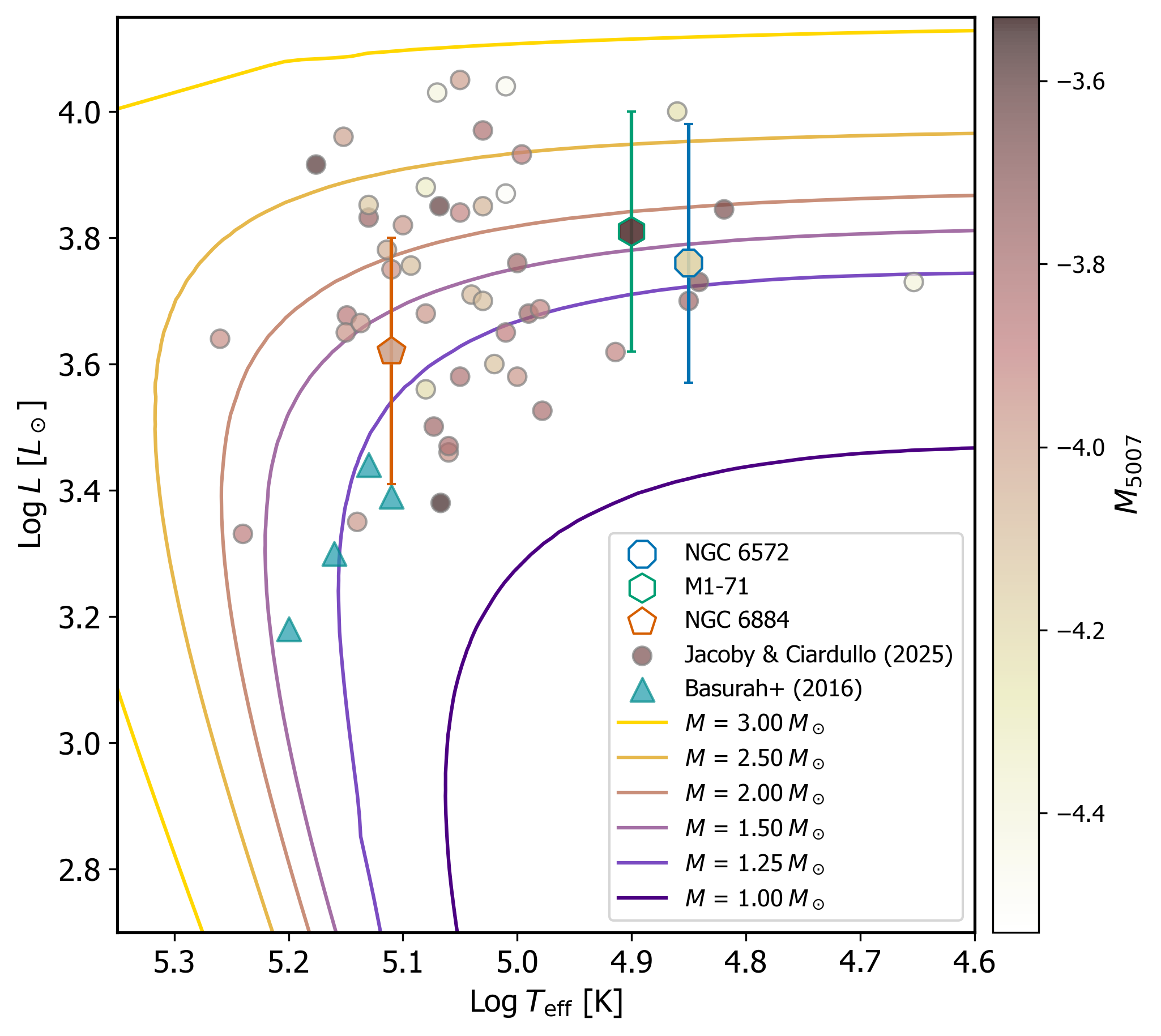}
   \caption{HR-diagram with the CSPN of the PNe in the top 1 mag of the PNLF from \citet{2025ApJ...983..129J}, which are indicated as circles with color gradient. The blue triangles mark the PN with \textit{wels} features from \citet{2016MNRAS.458.2694B}. The markers of our PNe and stellar tracks are the same as in Fig. \ref{fig:distance_tkin}.}
   \label{fig:HR_diagram_comparison}
\end{figure}

In addition to the excitation class, we also used the HR-diagram to infer the $M_{5007}$ phase, in comparison to the stellar tracks and the known \oiii-luminous PNe in the literature. \citet{2025ApJ...983..129J} have compiled the central star \teff~and $L_*$ of PNe within the top 1 mag of the PNLF in the LMC and M31. We compared those PNe with our sample in Fig. \ref{fig:HR_diagram_comparison}. Most of the \oiii-luminous PNe are found at the turn-off point of stellar tracks before going down into the white dwarf cooling tracks. \citet{2025ApJ...983..129J} inferred that PNe with $M_{5007} \lesssim-3.8$ have an average \oiii~conversion efficiency of $11.5 \pm2.2$\% (see their Fig. 3 for details). NGC\,6572, NGC\,6884, and M\,1-71 have the \oiii~conversion efficiencies of $12.8^{+7.8}_{-4.7}$\%, $14.9^{+7.7}_{-5.5}$\%, and $10.9^{+5.5}_{-3.8}$\%, respectively.

As discussed in Sect. \ref{sec:cs_class}, there is an ambiguity whether the \heii~emission line in NGC\,6572 and M\,1-71 comes from the star or from the nebula. However, based on the photoionisation models, we found that the $\mathrm{EC}_{\mathrm{low}}$ criteria is in a better agreement. The excitation class for NGC\,6572 and M\,1-71 is EC = 5.2 and EC = 5.1, respectively. This implies that both PNe are close to their peak $M_{5007}$ value. Moreover, if we see their position in the HR-diagram, they are still in the flat part of the stellar tracks. \citet{2007A&A...473..467S} suggested that at solar metallicity, the maximum $M_{5007}$ was achieved at $\sim 100 000$ K. While both PNe currently have lower \teff~values, we speculate that as the central stars get hotter in the future, the $M_{5007}$ will also get brighter; this might become a worthwhile subject for long-term monitoring. For NGC\,6572, \citet{2014ARep...58..702A} have recorded the increase of $F_{5007}$ between 1938 and 2013.

On the other hand, NGC\,6884 seems to be more evolved. From the stellar tracks, it may have turned already down into the white dwarf cooling sequence. We speculate that it has passed the maximum brightness of $M_{5007}$, as indicated from the excitation class of EC = 8.2 and \teff$> 100 000$ K. Since the mass of NGC\,6884 and NGC\,6572 is similar within their uncertainties, they may have a similar evolutionary path. They also within the mass range of the \textit{wels} PNe analysed by \citet{2016MNRAS.458.2694B}, which are likely more evolved than NGC\,6884; their PNe are further down into the cooling sequence with EC = 9 -- 10. 

\subsection{Implications for the PNLF cutoff} \label{sec:cspn_mass}

\begin{figure}[]
   \centering
   \includegraphics[width=\linewidth]{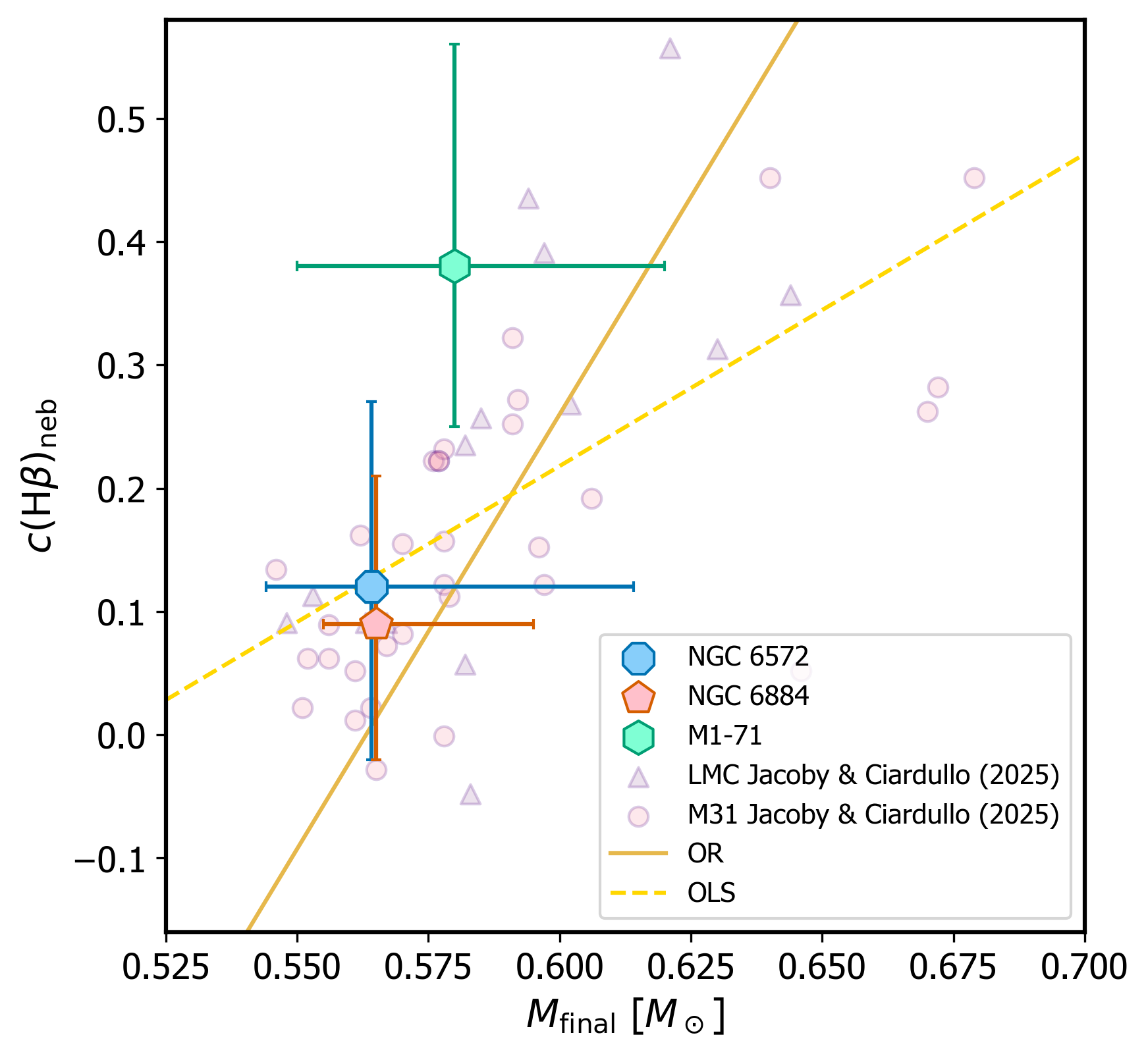}
   \caption{Empirical relation between central star final mass and circum-nebular extinction from \citep{2025ApJ...983..129J}. Our PNe are overplotted with the same markers as in Fig. \ref{fig:distance_tkin}.}
   \label{fig:extinction_vs_mass}
\end{figure}

We calculated the final mass ($M_\mathrm{fin}$) of the central stars based on the theoretically inferred $M_\mathrm{init}$ obtained from the stellar tracks of \citet[][MB16]{2016A&A...588A..25M}. Firstly, we derived the $M_\mathrm{fin}$ value using the initial-final mass relation (IFMR) from the same stellar tracks using the interpolated prescription by \citet{2025ApJ...983..129J}. Secondly, we adopted the IFMR from \citet[][C18]{2018ApJ...866...21C}, who empirically derived the relation from Sirius B and several cluster white dwarfs in the Galaxy. Thirdly, we also calculated the $M_\mathrm{fin}$ value using the empirically derived IFMR from the volume limited sample of Galactic white dwarfs within 40 pc \citep[][C24]{2024MNRAS.527.3602C}. The $M_\mathrm{fin}$ values are presented in Table \ref{tab:final_mass}. \citet{2025ApJ...983..129J} derived an empirical relation between $M_\mathrm{fin}$ and $c(\mathrm{H}\beta)_\mathrm{neb}$ for \oiii~luminous PNe in the LMC and M31. They defined the relation with two methods; orthogonal-regression (OR) and ordinary least square (OLS). Although the different methods yielded a different slope, they confirmed that the relation was real. They argued that the circum-nebular extinction plays a major role in the universality of the PNLF as a standard candle. We compare our derived circum-nebular extinction and $M_\mathrm{fin}$ to their trend in Fig. \ref{fig:extinction_vs_mass}. We only compare the $M_\mathrm{fin}$ from MB16 relation to be consistent with \citet{2025ApJ...983..129J}. Our PNe follow their general distribution within the uncertainties. 

\begin{table}
\caption{Final mass of our central star of PNe from different IFMR.}
\label{tab:final_mass}
\centering
\begin{tabular}{c c c c}
\hline
\hline
PN & MB16 [$M_\odot$]& C18 [$M_\odot$]& C24 [$M_\odot$] \\
\hline
\hline
NGC\,6572 & $0.56_{-0.02}^{+0.05}$ & $0.60_{-0.02}^{+0.09}$ & $0.59_{-0.02}^{+0.10}$\\
\hline
NGC\,6884 & $0.56_{-0.01}^{+0.03}$ & $0.60_{-0.01}^{+0.06}$ & $0.59_{-0.02}^{+0.06}$\\
\hline
M\,1-71 & $0.58_{-0.03}^{+0.04}$ & $0.63_{-0.04}^{+0.07}$ & $0.62_{-0.05}^{+0.08}$\\
\hline
\end{tabular}
\tablefoot{MB16: \citet{2016A&A...588A..25M}, C18: \citet{2018ApJ...866...21C}, C24: \citet{2024MNRAS.527.3602C}.
}
\end{table}

Our values showed that the MB16 relation gave the lowest $M_\mathrm{fin}$ compared to the values from the empirical IFMRs. Based on simulations, the maximum brightness of $M_{5007}$ is strongly correlated with the central star luminosity, and consequently the $M_\mathrm{fin}$ \citep{2007A&A...473..467S, 2018NatAs...2..580G, 2019ApJ...887...65V}. With the assumption that we needed at least $\sim$0.56 $M_\odot$ to achieve the PNLF cutoff magnitude $M_{*} = -4.53$ \citep{2019ApJ...887...65V, 2025A&A...699A.371V} at solar metallicity, our PNe should achieve this value and populate the PNLF cutoff within their lifetime, regardless of the adopted IFMR. Long-term monitoring and a larger PNe sample with similar properties would be necessary to confirm this. 

In a broader context, the comparison of $M_\mathrm{fin}$ from different IFMR has implications for understanding the universality of the PNLF cutoff. For example, in our sample, NGC\,6572 has the lowest $M_\mathrm{init}$ value. With $M_\mathrm{init} \sim 1.40 \: M_\odot$, it should have originated from a stellar population with the age of $\sim 2-4$ Gyr. If we adopt the MB16 IFMR, any older population will produce $M_\mathrm{fin} \lesssim 0.56$ and will fall short of the canonical $M_{*}$ value. However, if we adopted the C18 IFMR, a core mass of $\sim0.56 \:M_\odot$ can be obtained with  $M_\mathrm{init} \sim 1.0 \: M_\odot$; the progenitor population would have an age of $\sim 10$ Gyr. Stellar populations of this age dominate elliptical and lenticular galaxies, where the origin of the most \oiii~luminous PNe at the PNLF cutoff is questioned. This particular relationship between the IFMR and the PNLF cutoff has also been investigated using simulations with single stellar populations \citep[][see their Sect. 4.3 and Fig. 9 for details]{2025A&A...699A.371V}. Our findings support the argument that the IFMR plays a role in the universality of the PNLF cutoff across different stellar populations. 

\section{Conclusions and outlook} \label{sec:conclud}

We observed and analysed three PNe near to the Galactic PNLF cutoff. For each PN, we determined its nebular characteristics, chemical abundance, and central star parameters through photoionisation modelling. We also developed a method to constrain the PN distance, utilising the nebular \hb~luminosity ($L_{\mathrm{H}\beta}$), kinematical age ($t_{\mathrm{kin}}$), and age prediction from post-AGB stellar tracks ($t_{\mathrm{tracks}}$). We investigated the central star classification, progenitor mass, and its implication for the physical foundation of PNLF as a standard candle. We compared our sample with the most \oiii-luminous PNe in the Large Magellanic Cloud (LMC) and M31, emphasising the role of circum-nebular extinction in the PNLF cutoff. We also discussed the importance of the initial-to-final mass relation (IFMR) 
to explain the universality of PNLF standard candle across different stellar populations. Our conclusions are the following:

\begin{enumerate}

    \item Our distance estimates, the $L-t_{kin}$ distances, have errors comparable to those for the Gaia distances by \citet{2021A&A...656A.110C}. Adopting these values provide more consistent parameters in the HR-diagram, when compared to the theoretical predictions from the post-AGB models of \citet{2016A&A...588A..25M}. Based on these distances, we confirmed that our PNe are within 1 mag below the Galactic PNLF cutoff ($-4.20 \lesssim M_{5007} \lesssim-3.60$).
    
    \item Our PNe have similar nebular characteristics, with an electron temperature range of $T_e$[\ion{O}{iii}]$\,= 11350 - 11700$ K and an electron density range of log $N_e$[\ion{Ar}{iv}]$\,= 3.74 - 3.81$ cm$^{-2}$. They have chemical abundances that are typical for Galactic PNe within the uncertainties. NGC\,6572 and M\,1-71 have excitation classes of EC = 5.2 and EC = 5.1, respectively; this is theoretically expected in the most \oiii-luminous PNe and has been observed in nearby galaxies \citep{2007A&A...473..467S, 2022A&A...657A..71G, 2025ApJ...983..129J}. NGC\,6884 is slightly more evolved, with higher EC = 8.2, which was also confirmed by its position in the HR-diagram. The \oiii~conversion efficiencies of our PNe are in agreement with PNe at the PNLF cutoff of the LMC and M31.

    \item The measured extinction of our PNe is dominated by the foreground Galactic extinction. We emphasise the importance of only subtracting the foreground extinction to obtain the $M_{5007}$ to construct the Galactic PNLF. The circum-nebular extinction should be left uncorrected, as is the case in extragalactic PNLF studies where PNe are used as distance indicators and stellar population tracers. 

    \item The central stars in our sample exhibit \textit{wels} features; narrow emission lines of \ion{N}{iii}$\lambda4634-41$, \ion{C}{iii}$\lambda4647-50$, \ion{C}{iv}$\lambda4658$, \heii, and \ion{C}{iv}$\lambda5801-12$. We suggested that some of these lines likely arose from the nebula. In the on going debate of the \textit{wels} classification in the literature, it is more likely that our PNe have H-rich rather than H-poor central stars. 

    \item The progenitor stars of our PNe have initial masses range of $M_\mathrm{init} = 1.30-1.75 \, M_\odot$. The theoretical IFMR of  \citet{2016A&A...588A..25M} predicts a final mass range of $M_\mathrm{fin} = 0.56-0.58 \, M_\odot$. These values are lower than the ones derived from the empirical IFMR obtained from Galactic cluster white dwarf \citep{2018ApJ...866...21C}, which give a final mass range of $M_\mathrm{fin} = 0.60-0.63 \, M_\odot$. Future study on the IFMR is key to understand how the most \oiii-luminous PNe are produced in old stellar populations. 

\end{enumerate}

\noindent
This study demonstrates how we can study the most \oiii-luminous PNe in the Galaxy to understand the physical foundations behind the PNLF as a standard candle, with a future goal of comparing the extragalactic PNLF against the Galactic PNLF. This supports the effort to measure the Hubble constant ($H_0$) using the PNLF \citep{2021ApJ...916...21R, 2024ApJS..271...40J, 2025A&A...704A.303S}. A more extensive investigation with a larger sample of PNe at the Galactic PNLF cutoff is needed to determine whether our current findings are representative. Currently, our sample selection is based on the survey done by \citet{Chornay_Walton_Jones_Boffin_2023} that mostly covered northern Galactic PNe. Additional samples from the survey of the southern sky (Chornay et al., in prep) would be beneficial. It is particularly interesting to see, whether any of the most \oiii-luminous PNe in the Galaxy exhibit confirmed binary or H-poor central stars. This information would be crucial to improve existing PNLF models.

\begin{acknowledgements}
We thank the anonymous referee for their comments and suggestions, which helped improve the manuscript. The PMAS data are based on Guaranteed Observation Time (GTO) observations at the Centro Astronomico Hispano en Andalucia (CAHA) at Calar Alto, operated jointly by Junta de Andalucia and Cosejo Superior De Investigaciones Cientificias (IAA-CSIC). This research utilised the observations made with the NASA/ESA Hubble Space Telescope obtained from the Space Telescope Science Institute, which is operated by the Association of Universities for Research in Astronomy, Inc., under NASA contract NAS 5–26555. These observations are associated with the Program ID 9839 and 8390. This work made use of Astropy (http://www.astropy.org) a community-developed core Python package and an ecosystem of tools and resources for astronomy (Astropy Collaboration 2013, 2018, 2022). AS and MMR acknowledge support from DFG under grants RO 2213/40-1, RO 2213/41-1, RO 2213/42-1, RO 2213/43-1.     
\end{acknowledgements}

\bibliographystyle{aa}
\bibliography{pmas_pne}


\begin{appendix}

\section{Flux calibration precision} \label{app:flux_cal}

As flux standards, we observed standard stars BD+28D4211 (BD28), G191B2B (G191), and Feige 34 \citep{1990AJ.....99.1621O, 1990AJ.....99.1243T}. We derived sensitivity functions using the \textsc{P3D} software based on the reference spectra observed with the HST, which was accessible in the Calar Alto database\footnote{\url{https://www.caha.es/pedraz/SSS/HST_CALSPEC/hst_calspec.html}}. Based on each sensitivity function, we calculated the flux calibration precision using the equation below

\begin{equation}
    \mathrm{Precision [\%]} = \frac{F_{\mathrm{HST}} - F_{\mathrm{PMAS}}}{F_{\mathrm{HST}}} \times 100
\end{equation}

\begin{figure*}[h!]
   \centering
   \includegraphics[width=\linewidth]{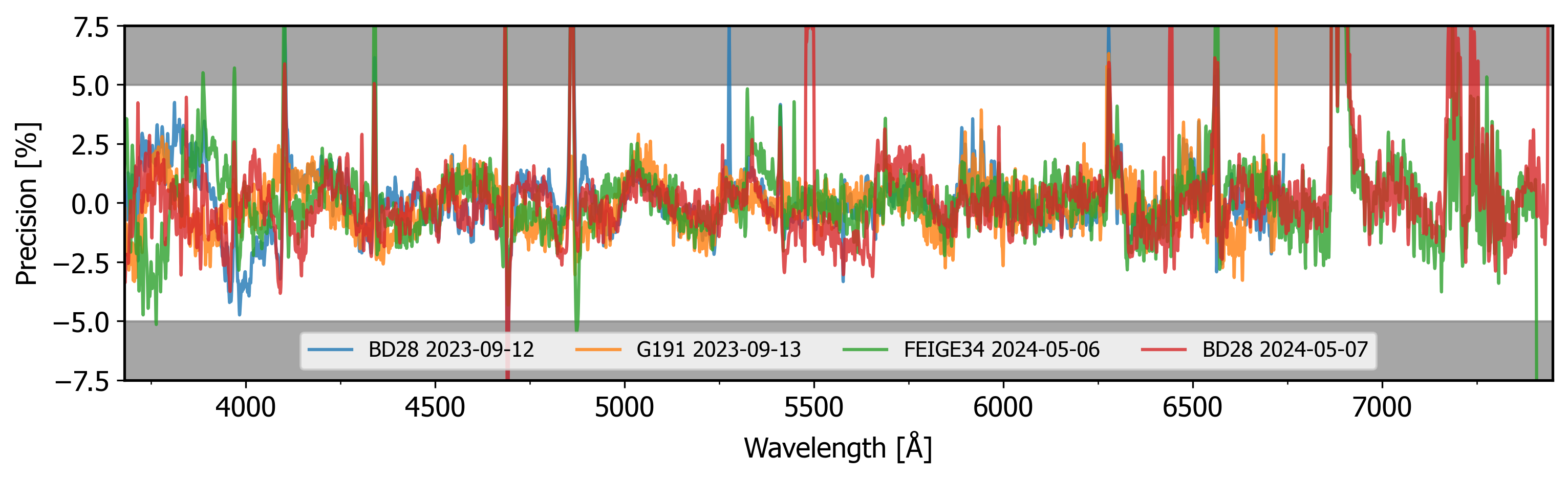}
   \caption{Flux calibration precision of four standard stars. Shaded regions indicate precision values more than $\pm 5\%$}.
   \label{fig:flux_cal}
\end{figure*}

\noindent
The flux calibration precision is plotted in Fig. \ref{fig:flux_cal}. 
Below 4000 \AA, they are consistent within $\pm 5\%$. At larger wavelengths, the precision of $2-3\%$ can be achieved. The visible spikes are caused either by bad pixels or cosmic rays in our observation. The features at $\sim6850-6950$\AA~and $\sim7150-7200$\AA~are atmospheric telluric lines. We did not fit any model to remove them and only used a sky subtraction to minimise their effect. Nevertheless, no emission lines from our PNe coincide with the telluric lines. For our analysis, we adopted $\pm 5\%$ flux calibration error throughout the wavelength range for consistency. We only employed BD28 as the main flux calibrator, as it was observed in every night in our programme. 

\section{Bad spaxel masking} \label{app:mosaic}

\begin{figure*}[h!]
   \centering
   \includegraphics[width=\linewidth]{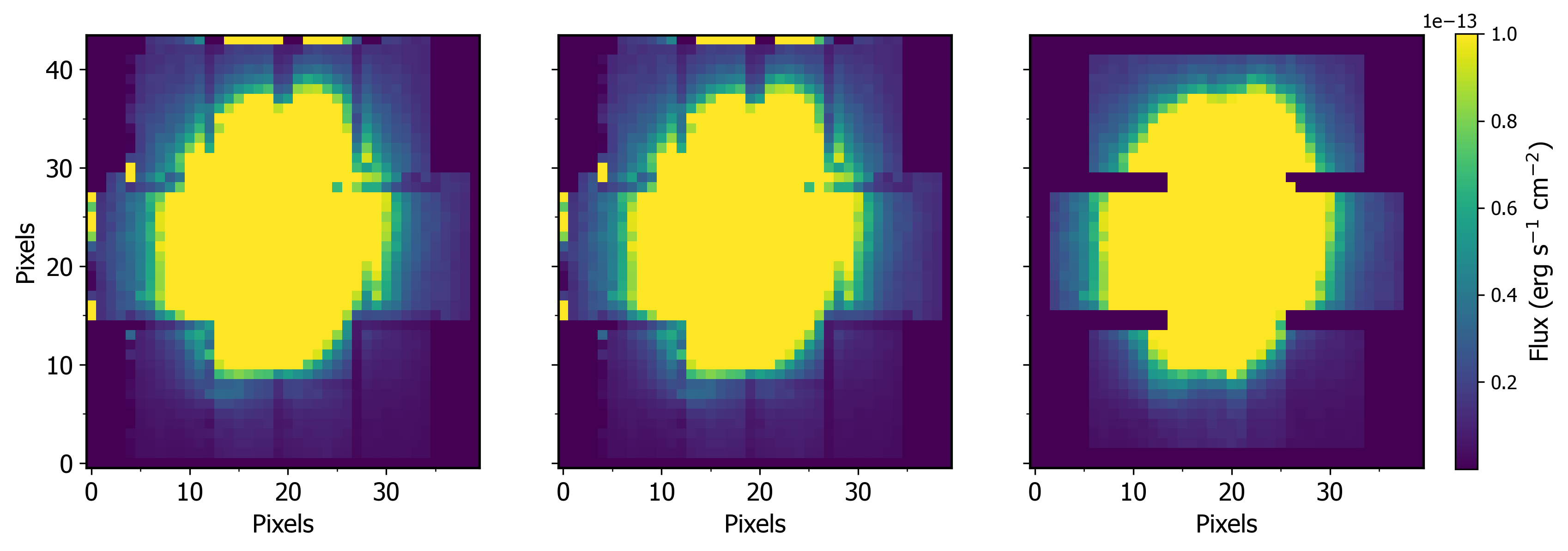}
   \caption{Emission line map of \oiii~line before any spaxel masking (left), after the sky exposure spaxel masking (middle), and after the exclusion of the outermost two rows and column (right).}.
   \label{fig:OIII_cleaning}
\end{figure*}

\begin{figure*}[h!]
   \centering
   \includegraphics[width=\linewidth]{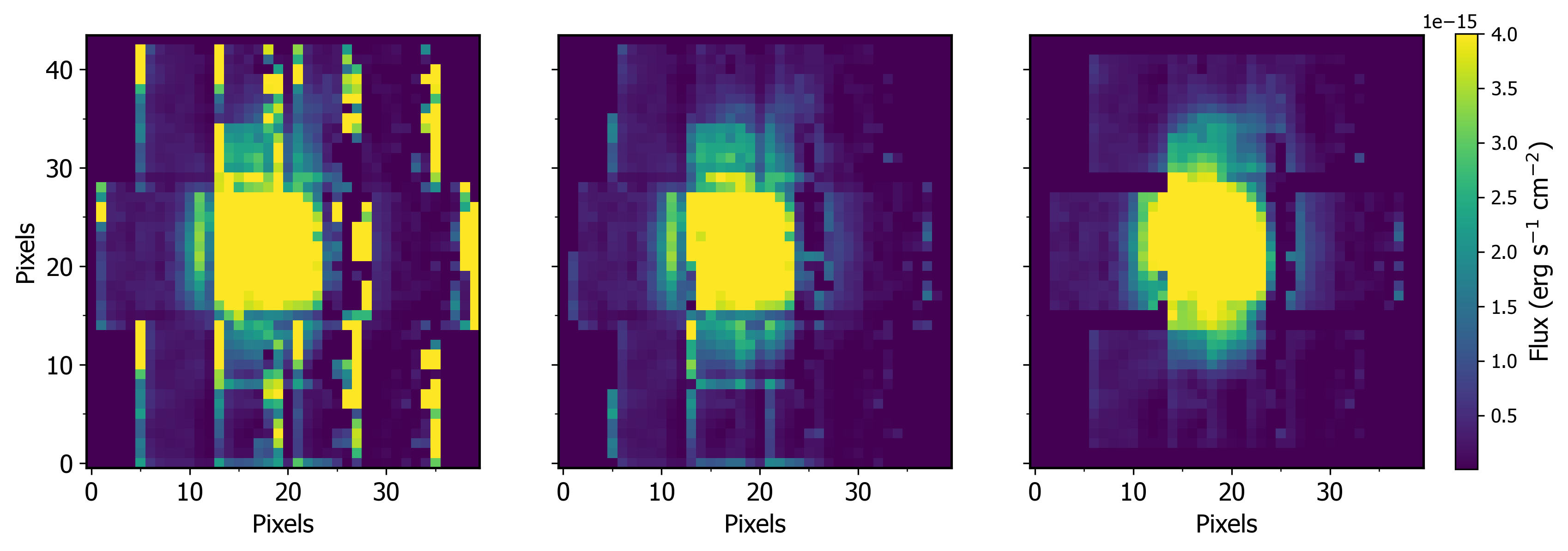}
   \caption{Same as Fig. \ref{fig:OIII_cleaning} but for the \sii~line.}.
   \label{fig:SII_cleaning}
\end{figure*}

As described in Sect. \ref{sec:data_red}, we encountered a vignetting and a possible fibre transmission degradation problem in our data. To demonstrate this, we used the data of NGC\,6572. Fig. \ref{fig:OIII_cleaning} and Fig. \ref{fig:SII_cleaning} show a sequence of images during the spaxel masking steps for the \oiii~and the \sii~line, respectively. In the unclean images on the left, some spaxels exhibit spurious values; this could falsely overestimate the integrated flux and in the particular case of NGC\,6572, affect the mosaicking procedure. The vignetting affected the outermost rows and columns for each individual pointing, and became worse at the edges of the wavelength range. This particularly affects \oii~and \sii~lines in the V600 setup and only \oii~line in V500 setup; there is no line of interest in the reddest part of the spectrum in V500 setting. We demonstrated this with the \sii~map of the NGC\,6572 data, which was observed with the V600 setup and appear close to the red spectrum edge in Fig. \ref{fig:SII_cleaning}. Furthermore, in some wavelengths, we could see similar behaving bad spaxels, but their position are not at the edges of the FOV; we speculated that it caused by the instrument age degradation. 

We found the position of these bad spaxels was consistent in the offset sky images, although their fluxes may vary. Firstly, we masked the bad spaxels in the sky exposure, which have flux larger than $5 \times 10^{-15}$ erg s$^{-1}$ cm$^{-2}$ \AA$^{-1}$; the result is the middle images. This was particularly effective to remove the spaxels caused by the speculated age degradation and partially effective against the vignetting effect. To further clean the image, we decided to mask the outermost two rows and columns of each pointing; the result is presented in the righten-most images. We applied this procedure for each pointing and and each object in our analysis. This process consequently caused some flux loss, which is discussed in the next appendix.   

\section{Flux loss correction} \label{app:flux_loss}

\begin{figure*}[h!]
   \centering
   \includegraphics[width=\linewidth]{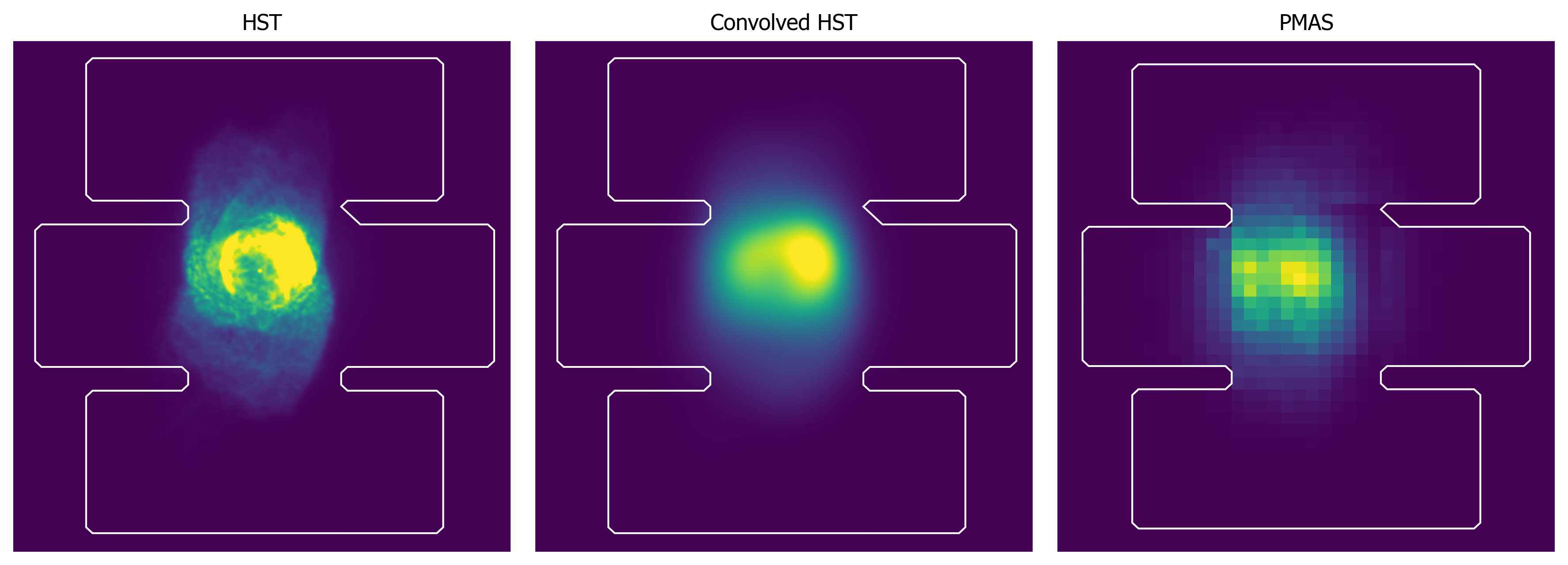}
   \caption{Comparison of images in the WFPC2-F502N filter for NGC\,6572 in the process of flux loss simulation. The left image is the original HST image. The middle image is the HST image convolved with the seeing FWHM of our observation. The right image is our PMAS observation.}.
   \label{fig:HST6572}
\end{figure*}

\begin{figure*}[h!]
   \centering
   \includegraphics[width=\linewidth]{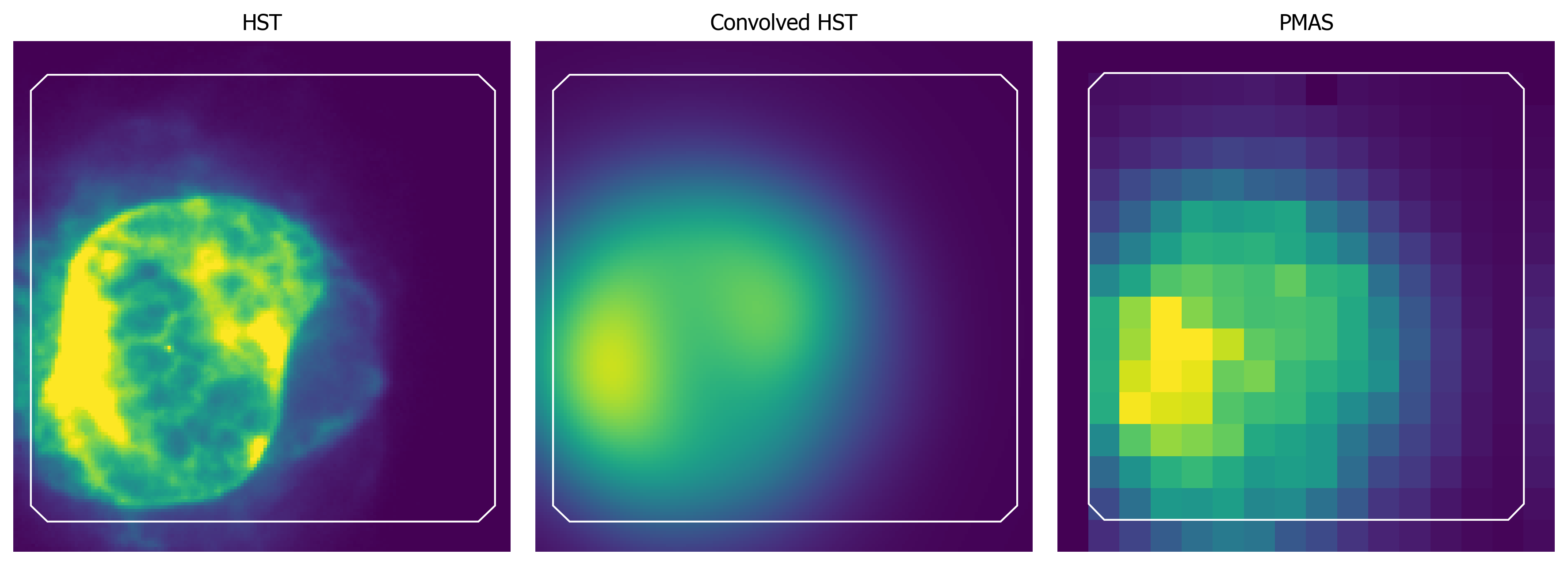}
   \caption{Same as Fig. \ref{fig:HST6572} but for NGC\,6884.}
   \label{fig:HST6884}
\end{figure*}

To make sure we capture the whole flux, especially to derive $M_{5007}$, we compared our images to the archival HST images obtained with the WFPC2 instrument and the F502N filter; they are available for NGC\,6572 (Program ID: 9839) and NGC\,6884 (Program ID: 8390). We convolved the HST images with our recorded seeing FWHM of 1\farcs5 for both objects. The comparison for NGC\,6572 and NGC\,6884 can be seen in Fig. \ref{fig:HST6572} and Fig. \ref{fig:HST6884}, respectively. As a comparison, we simulate our data using the bandpass of F502N filter. 

For the HST images, we constructed a curve of growth using a circular radius to determine the total flux. Then, we calculated the flux that was enclosed by the PMAS FOV and derived the flux loss correction factor. We estimated a flux loss of $\sim 1$\% and $\sim 17$\% for NGC\,6572 and NGC\,6884, respectively. We could not do this exact procedure for M\,1-71 as there was no HST image available.

\section{Atomic database} \label{app:line_emissivity}

Atomic data references that we utilised to derive the ionic abundances are tabulated in Table \ref{tab:atomic_ref}. These are included in the default \textsc{PyNeb} database.  

\begin{table*}
\caption{References for different parameters that were employed for determining the of ionic abundances.}
\label{tab:atomic_ref}
\centering
\begin{tabular}{c c c c}
\hline
\hline
Atom / Ion & $A$-levels & Energy levels & Collisional strengths \\
\hline
\hline
H & \citet{1995MNRAS.272...41S} & \citet{2010ADNDT..96..586K} & --\\
\hline
He & \citet{2012MNRAS.425L..28P, 2013MNRAS.433L..89P} & -- & --\\
\hline
N$^{+}$ & \citet{2004ADNDT..87....1F} & \citet{1993tshc.book.....M} & \citet{2011ApJS..195...12T}\\
\hline
O$^{0}$ & \citet{1996atpc.book.....W} & \citet{1993tshc.book.....M} & \citet{1995ApJS...96..325B}\\
O$^{+}$ & \citet{1982MNRAS.198..111Z, 1996atpc.book.....W} & \citet{1993JPCRD..22.1179M} & \citet{2009MNRAS.397..903K}\\
O$^{++}$ & \citet{2000MNRAS.312..813S, 2004ADNDT..87....1F} & \citet{1993tshc.book.....M} & \citet{2014MNRAS.441.3028S}\\
\hline
Ne$^{++}$ & \citet{Galavis2019} & \citet{2006EPJD...37....1K} & \citet{2011JPhB...44q5206M}\\
\hline
S$^{+}$ & \citet{Rynkun2019} & \citet{1990JPCRD..19..821M} & \citet{Tayal2010}\\
S$^{++}$ & \citet{2006ADNDT..92..607F} & \citet{1990JPCRD..19..821M} & \citet{1999ApJ...526..544T}\\
\hline
Cl$^{++}$ & \citet{Rynkun2019} & \citet{1934PhRv...45..401B} & \citet{Butler1989}\\
\hline
Ar$^{++}$ & \citet{MunosBurgos2009} & \citet{2010JPCRD..39c3101S} & \citet{MunosBurgos2009}\\
Ar$^{+++}$ & \citet{Rynkun2019} & \citet{2010JPCRD..39c3101S} & \citet{1997MNRAS.284..754R}\\
\hline
\end{tabular}
\end{table*}

\section{Line intensity and model comparison} \label{app:line_flux}

The line intensities of our observation and photoionisation model for NGC\,6572, NGC\,6884, and M\,1-71 are presented in Table \ref{tab:data_table_NGC6572}, Table \ref{tab:data_table_NGC6884}, and Table \ref{tab:data_table_M171}, respectively. 

\begin{table*}[]
\caption{NGC\,6572 line FWHM, intensities, and model comparison.}
\label{tab:data_table_NGC6572}
\centering
\begin{tabular}{c | c c | c c | c}
\hline
\hline
Parameter & \multicolumn{4}{c}{Observation} & Model \\

\hline
$I(\mathrm{H}\beta)$ [$\mathrm{erg \,cm^{-2}\,s^{-1}}$]            &  \multicolumn{4}{c}{$4.72_{-1.24}^{+1.96} \times10^{-10}$}             &  - \\
$L(\mathrm{H}\beta)$ [$\mathrm{erg \,s^{-1}}$]            &  \multicolumn{4}{c}{$2.00_{-0.72}^{+1.34} \times10^{35}$}             &  $1.93\times10^{35}$\\
$T_e$[\ion{O}{iii}] [K] & \multicolumn{4}{c}{11400$_{-250}^{+290}$} & 11438 \\
log $L_{\mathrm{CSPN}}$ & \multicolumn{4}{c}{3.76$_{-0.19}^{+0.22}$} & 3.76 \\
log \teff$_{,\mathrm{CSPN}}$ & \multicolumn{4}{c}{-} & 4.85 \\
\hline
Line & FWHM [Å]& $\Delta$FWHM [Å] & $I(\lambda)$ & $\Delta I(\lambda)$ & $I(\lambda)$ \\
\hline
$[$\ion{O}{ii}$]\lambda3727,29$ & 4.88 & 0.19 & 50.41 & 6.17 & 67.07 \\
$[$\ion{Ne}{iii}$]\lambda3869$ & 3.11 & 0.03 & 118.67 & 13.74 & 116.05 \\
$[$\ion{O}{iii}$]\lambda4363$ & 3.20 & 0.03 & 11.60 & 0.86 & 12.26 \\
\ion{He}{ii}$\, \lambda4686$\tablefootmark{a} & 4.30 & 0.24 & 0.91 & 0.05 & 0.02 \\
$[$\ion{Ar}{iv}$]\lambda4712$\tablefootmark{b} & 3.95 & 0.07 & 2.41 & 0.13 & 2.56 \\
$[$\ion{Ar}{iv}$]\lambda4740$ & 3.18 & 0.04 & 2.84 & 0.15 & 2.22 \\
\ion{H}{i}$\, \lambda4861$ & 2.77 & 0.02 & 100.00 & 0.48 & 100.00 \\
$[$\ion{O}{iii}$]\lambda5007$ & 2.57 & 0.01 & 1147.06 & 57.43 & 1170.98 \\
$[$\ion{Cl}{iii}$]\lambda5518$ & 3.28& 0.32 & 0.34 & 0.02 & 0.40 \\
$[$\ion{Cl}{iii}$]\lambda5538$ & 3.42& 0.11& 0.73 & 0.05 & 0.75 \\
$[$\ion{N}{ii}$]\lambda5575$ & 3.08& 0.03& 3.03 & 0.25 & 1.81 \\
\ion{C}{iv}$\lambda5801$\tablefootmark{a} & 6.08 & 1.18 & 0.30 & 0.05 & -- \\
\ion{He}{i}$\, \lambda5876$& 2.74 & 0.03 & 17.46 & 1.51 & 16.43 \\
$[$\ion{O}{i}$]\lambda6300$ & 3.07 & 0.02 & 8.45 & 0.90 & 3.85 \\
$[$\ion{O}{i}$]\lambda6363$ & 3.26 & 0.04 & 2.95 & 0.33 & 1.23 \\
\ion{H}{i}$\, \lambda6563$ & 2.70 & 0.11 & 282.48 & 8.28 & 277.83 \\
$[$\ion{N}{ii}$]\lambda6583$ & 2.82 & 0.04 & 78.12 & 8.92 & 70.82 \\
\ion{He}{i}$\, \lambda6678$ & 3.08 & 0.03 & 4.94 & 0.59 & 4.23 \\
$[$\ion{S}{ii}$]\lambda6716$ & 3.03& 0.10 & 1.21 & 0.14 & 1.43 \\
$[$\ion{S}{ii}$]\lambda6731$ & 3.20& 0.47& 2.71 & 0.33 & 2.80 \\
\hline
\end{tabular}
\tablefoot{
\tablefoottext{a}{Possible central star origin. \ion{C}{iv}$\lambda5801$ line is not calculated in our photoionisation model.}
\tablefoottext{b}{Possible blend with \ion{He}{I}$\lambda4713$ line.}
}
\end{table*}

\begin{table*}[]
\caption{NGC\,6884 line FWHM, intensities, and model comparison.}
\label{tab:data_table_NGC6884}
\centering
\begin{tabular}{c | c c | c c | c}
\hline
\hline
Parameter & \multicolumn{4}{c}{Observation} & Model \\

\hline
$I(\mathrm{H}\beta)$ [$\mathrm{erg \,cm^{-2}\,s^{-1}}$]            &  \multicolumn{4}{c}{$3.96_{-0.92}^{+1.21} \times10^{-11}$}             & -\\
$L(\mathrm{H}\beta)$ [$\mathrm{erg \,s^{-1}}$]            &  \multicolumn{4}{c}{$1.43_{-0.55}^{+0.72} \times10^{35}$}             &  $1.32\times10^{35}$\\
$T_e$[\ion{O}{iii}] [K] & \multicolumn{4}{c}{11680$_{-220}^{+210}$} & 12157 \\
log $L_{\mathrm{CSPN}}$ & \multicolumn{4}{c}{3.62$_{-0.21}^{+0.18}$} & 3.62 \\
log \teff$_{,\mathrm{CSPN}}$ & \multicolumn{4}{c}{-} & 5.11 \\
\hline
Line & FWHM [Å]& $\Delta$FWHM [Å] & $I(\lambda)$ & $\Delta I(\lambda)$ & $I(\lambda)$ \\
\hline
$[$\ion{O}{ii}$]\lambda3727,29$ & 5.66 & 0.37 & 29.48 & 2.91 & 126.99 \\
$[$\ion{Ne}{iii}$]\lambda3869$ & 3.81 & 0.03& 148.13 & 13.18 & 150.36 \\
$[$\ion{O}{iii}$]\lambda4363$ & 4.10 & 0.09 & 14.45 & 0.94 & 16.62 \\
\ion{He}{ii}$\, \lambda4686$ & 3.92 & 0.02& 23.12 & 1.21 & 14.88 \\
$[$\ion{Ar}{iv}$]\lambda4712$\tablefootmark{a} & 4.46 & 0.06& 5.9 & 0.31 & 4.64 \\
$[$\ion{Ar}{iv}$]\lambda4740$ & 4.02 & 0.06& 6.76 & 0.35 & 3.68 \\
\ion{H}{i}$\, \lambda4861$ & 3.63 & 0.03 & 100.0 & 0.69 & 100.0 \\
$[$\ion{O}{iii}$]\lambda5007$ & 3.32 & 0.02 & 1344.61 & 65.78 & 1377.18 \\
$[$\ion{Cl}{iii}$]\lambda5518$ & 4.05 & 0.41& 0.59 & 0.04 & 0.65 \\
$[$\ion{Cl}{iii}$]\lambda5538$ & 4.17 & 0.24& 0.84 & 0.05 & 0.88 \\
$[$\ion{N}{ii}$]\lambda5575$ & 4.32 & 0.10 & 1.19 & 0.08 & 4.26 \\
\ion{C}{iv}$\lambda5801$ & -- & -- & -- & -- & -- \\
\ion{He}{i}$\, \lambda5876$& 3.60 & 0.03& 16.24 & 1.16 & 15.59 \\
$[$\ion{O}{i}$]\lambda6300$ & 3.65 & 0.54 & 2.6 & 0.23 & 6.52 \\
$[$\ion{O}{i}$]\lambda6363$ & 4.02 & 0.45 & 0.99 & 0.08 & 2.08 \\
\ion{H}{i}$\, \lambda6563$ & 3.01 & 0.04 & 283.18 & 3.17 & 277.97 \\
$[$\ion{N}{ii}$]\lambda6583$ & 3.38  & 0.04 & 40.2 & 3.82 & 183.82 \\
\ion{He}{i}$\, \lambda6678$ & 3.84 & 0.08& 4.57 & 0.44 & 12.04 \\
$[$\ion{S}{ii}$]\lambda6716$ & 3.80 & 0.22 & 1.8 & 0.17 & 4.42 \\
$[$\ion{S}{ii}$]\lambda6731$ & 3.72 & 0.49 & 3.33 & 0.32 & 7.62 \\
$[$\ion{Ar}{ii}$]\lambda7135$ & 3.42 & 0.02& 16.21 & 1.78 & 18.64  \\
$[$\ion{O}{ii}$]\lambda7220$ & 3.93 & 1.01 & 1.98 & 0.24 & 6.88 \\
$[$\ion{O}{ii}$]\lambda7230$ & 3.90 & 0.89 & 1.68 & 0.19 & 5.63 \\
\hline
\end{tabular}
\tablefoot{
\tablefoottext{a}{Possible blend with \ion{He}{I}$\lambda4713$ line.}
}
\end{table*}

\begin{table*}[]
\caption{M\,1-71 line FWHM, intensities, and model comparison.}
\label{tab:data_table_M171}
\centering
\begin{tabular}{c | c c | c c | c}
\hline
\hline
Parameter & \multicolumn{4}{c}{Observation} & Model \\

\hline
$I(\mathrm{H}\beta)$ [$\mathrm{erg \,cm^{-2}\,s^{-1}}$]            &  \multicolumn{4}{c}{$9.81_{-2.50}^{+2.98} \times10^{-11}$}             & -\\
$L(\mathrm{H}\beta)$ [$\mathrm{erg \,s^{-1}}$]            &  \multicolumn{4}{c}{$2.22_{-0.76}^{+1.25} \times10^{35}$}             &  $2.18\times10^{35}$\\
$T_e$[\ion{O}{iii}] [K] & \multicolumn{4}{c}{11350$_{-280}^{+310}$} & 11737 \\
log $L_{\mathrm{CSPN}}$ & \multicolumn{4}{c}{3.81$_{-0.19}^{+0.19}$} & 3.81 \\
log \teff$_{,\mathrm{CSPN}}$ & \multicolumn{4}{c}{-} & 4.90 \\
\hline
Line & FWHM [Å]& $\Delta$FWHM [Å] & $I(\lambda)$ & $\Delta I(\lambda)$ & $I(\lambda)$ \\
\hline
$[$\ion{O}{ii}$]\lambda3727,29$ & 4.94 & 0.69 & 56.25 & 5.83 & 70.81 \\
$[$\ion{Ne}{iii}$]\lambda3869$ & 4.09 & 0.05 & 195.14 & 19.04 & 214.49 \\
$[$\ion{O}{iii}$]\lambda4363$ & 3.94 & 0.24 & 11.15 & 0.73 & 12.79 \\
\ion{He}{ii}$\, \lambda4686$ & -- & -- & -- & -- & 0.10 \\
$[$\ion{Ar}{iv}$]\lambda4712$\tablefootmark{a} & 4.26 & 0.60 & 2.06 & 0.1 & 2.63 \\
$[$\ion{Ar}{iv}$]\lambda4740$ & 3.65 & 0.42 & 2.71 & 0.14 & 2.18 \\
\ion{H}{i}$\, \lambda4861$ & 3.56 & 0.03 & 100.0 & 0.61 & 100.0 \\
$[$\ion{O}{iii}$]\lambda5007$ & 3.16 & 0.02 & 1120.05 & 56.45 & 1156.24 \\
$[$\ion{Cl}{iii}$]\lambda5518$ & -- & -- & -- & -- & 0.62 \\
$[$\ion{Cl}{iii}$]\lambda5538$ & 3.61 & 0.38& 0.64 & 0.04 & 1.01 \\
$[$\ion{N}{ii}$]\lambda5575$ & 3.86 & 0.15 & 3.87 & 0.28 & 2.70 \\
\ion{C}{iv}$\lambda5801$ & -- & -- & -- & -- & -- \\
\ion{He}{i}$\, \lambda5876$& 3.30 & 0.05 & 19.4 & 1.45 & 17.54 \\
$[$\ion{O}{i}$]\lambda6300$ & 3.42 & 0.15 & 8.84 & 0.79 & 3.64 \\
$[$\ion{O}{i}$]\lambda6363$ & 3.54 & 0.05 & 3.07 & 0.27 & 1.16 \\
\ion{H}{i}$\, \lambda6563$ & 2.91 & 0.05& 283.03 & 2.8 & 277.78 \\
$[$\ion{N}{ii}$]\lambda6583$ & 2.89 & 0.03 & 87.48 & 8.12 & 110.30 \\
\ion{He}{i}$\, \lambda6678$ & 3.36 & 0.03 & 5.01 & 0.51 & 5.87 \\
$[$\ion{S}{ii}$]\lambda6716$ & 3.55 & 0.25 & 1.38 & 0.13 & 2.05 \\
$[$\ion{S}{ii}$]\lambda6731$ & 3.64 & 0.38 & 2.76 & 0.28 & 3.82 \\
$[$\ion{Ar}{ii}$]\lambda7135$ & 3.17 & 0.03 & 17.27 & 2.02 & 19.36  \\
$[$\ion{O}{ii}$]\lambda7220$ & 3.39 & 0.71 & 6.92 & 0.86 & 5.30 \\
$[$\ion{O}{ii}$]\lambda7230$ & 3.69 & 0.16 & 5.82 & 0.7 & 4.35 \\
\hline
\end{tabular}
\tablefoot{
\tablefoottext{a}{Possible blend with \ion{He}{I}$\lambda4713$ line.}
}
\end{table*}

\end{appendix}

\end{document}